\documentclass[review, fleqn, preprint, 3p, 12pt]{elsarticle}

\usepackage{picinpar}
\usepackage{url}
\usepackage{flushend}
\usepackage{colortbl}
\usepackage{soul}
\usepackage{multirow}
\usepackage{pifont}
\usepackage{color}
\usepackage{alltt}
\usepackage{enumerate}
\usepackage{epstopdf}
\usepackage{pbox}
\usepackage{array}
\usepackage{nomencl}

\usepackage{booktabs}
\usepackage[T1]{fontenc}
\usepackage{setspace}

\usepackage[tight]{subfigure}
\usepackage{amsmath,amssymb,amscd,graphicx, color}
\usepackage{adjustbox}

\allowdisplaybreaks

\newcommand*{\red}{\textcolor{black}}

\begin{document}


\begin{frontmatter}

\title{Model-Free Reconstruction of Capacity Degradation Trajectory of Lithium-Ion Batteries Using Early Cycle Data}

\author[mymainaddress]{Seongyoon Kim\fnref{cor1}}
\author[mymainaddress]{Hangsoon Jung\fnref{cor1}}
\author[mymainaddress]{Minho Lee}
\author[mymainaddress]{Yun Young Choi}
\author[mymainaddress]{Jung-Il Choi\corref{cor1}}\ead{jic@yonsei.ac.kr}
\address[mymainaddress]{School of Mathematics and Computing (Computational Science and Engineering), Yonsei University, Seoul 03722, Republic of Korea}
\fntext[cor1]{These authors contributed equally to this work and should be considered co-first authors.}
\cortext[cor1]{Corresponding author.}

\begin{abstract}

Early degradation prediction of lithium-ion batteries is crucial for ensuring safety and preventing unexpected failure in manufacturing and diagnostic processes. Long-term capacity trajectory predictions can fail due to cumulative errors and noise. To address this issue, this study proposes a data-centric method that uses early single-cycle data to predict the capacity degradation trajectory of lithium-ion cells. The method involves predicting a few knots at specific retention levels using a deep learning-based model and interpolating them to reconstruct the trajectory. Two approaches are used to identify the retention levels of two to four knots: uniformly dividing the retention up to the end of life and finding optimal locations using Bayesian optimization.
The proposed model is validated with experimental data from 169 cells using five-fold cross-validation. The results show that mean absolute percentage errors in trajectory prediction are less than 1.60\% for all cases of knots. By predicting only the cycle numbers of at least two knots based on early single-cycle charge and discharge data, the model can directly estimate the overall capacity degradation trajectory.
Further experiments suggest using three-cycle input data to achieve robust and efficient predictions, even in the presence of noise. The method is then applied to predict various shapes of capacity degradation patterns using additional experimental data from 82 cells. The study demonstrates that collecting only the cycle information of a few knots during model training and a few early cycle data points for predictions is sufficient for predicting capacity degradation. This can help establish appropriate warranties or replacement cycles in battery manufacturing and diagnosis processes.
\end{abstract}


\end{frontmatter}

\section{Introduction}

Lithium-ion batteries (LIBs) are prominent power sources that are used in applications across diverse industries.
Factors such as high energy density, long lifespan, and low self-discharge make them suitable for applications in electric vehicles (EVs) and energy storage systems~\cite{armand2008building, dunn2011electrical, choi2021parameter}.
However, the batteries undergo degradation due to various causes, such as operational and environmental conditions, and this reduces the capacity and increases the internal resistance, leading to degradation of the available energy and power~\cite{barre2013review, birkl2017degradation}.

Battery degradation involves multiple complex physical mechanisms depending on the chemistry, usage conditions, and surrounding environment.
The state of health (SOH) and remaining useful life (RUL) are the primary indicators of battery health.
The SOH represents the ratio of the current capacity to the nominal (or initial) capacity and gradually decreases with degradation.
The RUL represents the remaining number of charge and discharge cycles till the end of life (EOL).
Hence, accurate prediction of battery SOH and RUL is required to inform the user whether a battery should be replaced and to avoid unexpected capacity fade~\cite{zhang2020identifying}.
Meanwhile, the capacity degradation patterns have a two-phase characteristic that is initially slow but suddenly accelerates after a transition point.
This pattern is referred to as the knee pattern, based on the shape of the capacity curve, and its transition point is called the knee point~\cite{schuster2015nonlinear, attia2022knees}.
Unlike SOH and RUL predictions, the knee point prediction allows early detection of rapid decrease, leading to more effective predictive maintenance~\cite{fermin2020identification}.
The prediction of knee point is beneficial for manufacturers that are responsible for the processing and diagnosis of batteries, and thus, the remaining cycle numbers upon knee points have been included in the main indicators to be predicted in addition to SOH and RUL.


Driven by the flexibility in applications without complex physical models, data-driven approaches are attracting considerable attention from both the academic and industrial fields~\cite{lyu2019model, zhou2020remaining, kim2021forecasting}.
Recently, machine learning with a feature-based approach achieved early predictions of RUL and knee points with high accuracy~\cite{fermin2020identification, severson2019data, shu2021state}.
Machine-learning-based approaches involve manual feature extraction to enhance the model's predictive performance.
For example, the knee point prediction
accuracy using relevance vector machine achieved 9.4\% mean absolute percentage error using the first 50 cycle information of the cells~\cite{fermin2020identification}.
In contrast, deep learning methods based on neural networks have been increasingly used owing to their automatic feature extraction and excellent generalization ability~\cite{kong2022state, hsu2022deep, kim2022impedance}.
In particular, because batteries have complex degradation mechanisms and generate massive voltage and current measurements, the convolutional neural network (CNN)~\cite{bengio2009learning} is widely used owing to benefits such as excellent feature extraction properties due to sparse interactions and parameter sharing~\cite{shen2020deep, li2021lithium}.
Strange and Dos Reis~\cite{strange2021prediction} used individual CNN models to predict the knee onset, knee point, and EOL using only one preliminary cycle.

Although these data-driven approaches can predict the RUL and knee point with good accuracy, some issues still persist.
A majority of the studies have focused on point prediction of the RUL or knee point, and only few have performed long capacity trajectory prediction.
The reason is assumed to be that the prediction model has to make thousands of extrapolation predictions for the long capacity trajectory, leading to prediction failure due to cumulative errors and noise~\cite{kim2021forecasting, li2020state}.
Instead of predicting all capacities on the trajectory, Herring \textit{et al}.~\cite{herring2020beep} developed models to predict the cycle numbers of a few points as a function of capacity, which further decreases the model dimension.
Strange and Dos Reis~\cite{strange2021prediction} reconstructed the capacity and internal resistance trajectory from predicted knee onset, knee point, and EOL using a simple empirical model.
Saxena \textit{et al}.~\cite{saxena2022convolutional} proposed a critical indicator of battery performance degradation in a four-parameter bilinear equation to reconstruct the capacity trajectory.
Deng \textit{et al}.~\cite{deng2022battery} classified the degradation patterns and used long short-term memory network with transfer learning for efficient trajectory predictions.
However, as knee point involves models such as the Bacon and Watts (BW) model~\cite{bacon1971estimating}, a discrepancy in the capacity degradation trajectory may occur for other trajectory shapes.
Therefore, it is necessary to develop a model-free methodology that is not constrained by the model when reconstructing a trajectory for general applicability to various shapes of battery degradation patterns.



This paper proposes a novel model-free method to reconstruct the capacity degradation trajectory of LIB cells by predicting a few knots, including the RUL based on early single-cycle data.
In general, the capacity degradation trajectory of a battery decreases monotonically owing to various degradation mechanisms.
Therefore, given the cycle numbers of a few knots at specific SOH levels, the capacity trajectory can be well approximated by interpolating the knots.
The main difference between the proposed method and the existing methods lies in the model output.
The proposed model predicts intervals in cycle numbers of knots for the given SOH levels rather than the capacity values of knots.
\red{
It should be noted that the knee point can be an example of the knot to express the capacity degradation pattern, where the battery capacity degrades gradually and accelerates from the knee point.}
In this study, two approaches are considered to select the SOH levels of knots: uniformly divide the SOHs up to EOL and find the optimal locations using Bayesian optimization~\cite{snoek2012practical}, which yield the best reconstruction of the capacity trajectory.
The selected knots are referred to as uniform and optimized knots, respectively.
The proposed method was validated using experimental data from 169 LIB cells generated by Severson \textit{et al}.~\cite{severson2019data} and Attia \textit{et al}.~\cite{attia2020closed} and applied to predict various degradation patterns of 82 cells generated by Zhu \textit{et al}.~\cite{zhu2022data}.
The proposed model-free approach can yield better generalization of the capacity trajectory reconstruction regardless of the battery degradation patterns, and simultaneous predictions of multiple knots can alleviate cumulative errors in the long-term capacity trajectory predictions.
The novelty of this study can be highlighted by the model-free reconstruction of the capacity degradation trajectory for LIBs using the predicted knots, regardless of the shapes of trajectories.
In addition, a quantitative analysis was conducted on the impact of the number of input cycles on capacity trajectory, which can help determine the number of input cycles in future studies.
We demonstrate that the proposed model can directly estimate the overall trajectory by predicting only the cycle numbers of at least two knots based on a few early cycle charge and discharge data.

The remainder of this paper is organized as follows.
The data and model structure, along with detailed implementations, are presented in Section~\ref{sec2:method}, and Section~\ref{sec3:results} presents the simulation results and discussion.
Finally, Section~\ref{sec4:conclusion} presents the major conclusions drawn from this study and outlines the scope of future research.


\section{Methodology and Implementation}\label{sec2:method}

\begin{figure}[t]
    \centering
    \includegraphics[width=\textwidth]{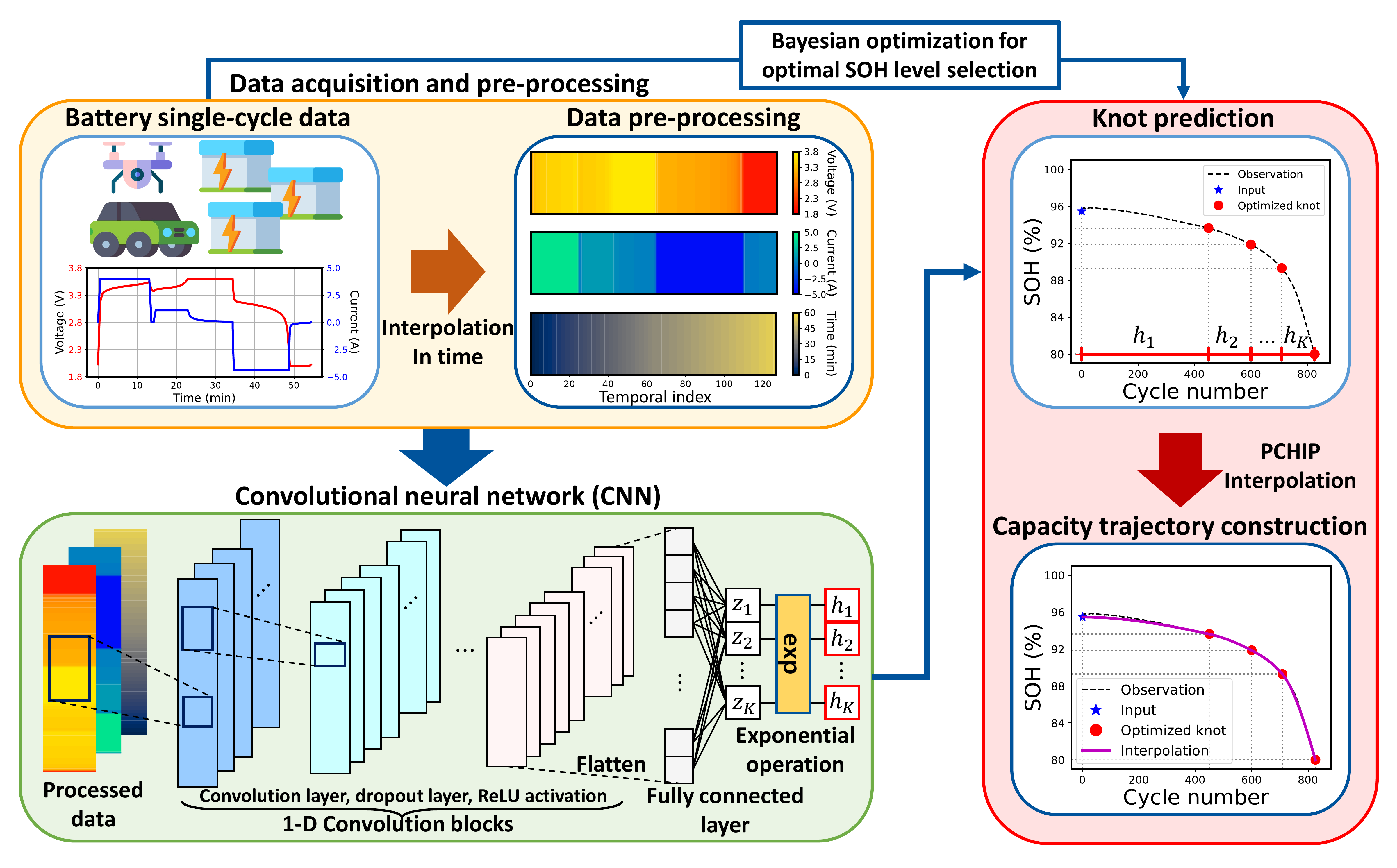}
    \caption{Framework of the proposed method.}
    \label{fig:libschema}
\end{figure}

The overall framework of the proposed method is illustrated in Fig.~\ref{fig:libschema}.
In the data acquisition and pre-processing part, a sequential data processing method was applied to raw data for the voltage, current, and corresponding time, in order to ensure equal dimensions in the data.
The pre-processed data were then applied to the CNN model, which automatically extracts features from the data.
The model predicts the intervals in cycle numbers of knots in specific SOH levels, which are determined uniformly or using Bayesian optimization.
Finally, the predicted locations of knots are used to reconstruct the capacity degradation trajectory using the interpolation.
The detailed process of the proposed method is described as follows.

\subsection{Data pre-processing}\label{sec2.1:preprocessing}

To construct the overall capacity degradation trajectory, the knots are selected on the trajectory.
The knots contain a cycle number of specific SOH values including EOL (e.g., values between 80 and 95\%).
The capacity degradation trajectory with a monotonically decreasing pattern can be approximated by applying piecewise cubic Hermite interpolating polynomial (PCHIP) interpolation~\cite{fritsch1984method} on the knots.
The PCHIP interpolation preserves monotonicity in the interpolation data and does not overshoot if the data is not smooth.
In PCHIP interpolation, when interpolating for a given variable $y_k$ at cycle $p_k$, the first derivatives at the points $p_k$, $f_k'$ are used.
Letting $h_k=p_{k+1}-p_k$ and $d_k=(y_{k+1}-y_k)/h_k$, then $f_k'$ is calculated by the weighted harmonic mean as follows.
\begin{align}
    \frac{w_1 + w_2}{f_k'} = \frac{w_1}{d_{k-1}} + \frac{w_2}{d_k},
\end{align}
where $w_1$ and $w_2$ equal $2h_k+h_{k-1}$ and $h_k + 2h_{k-2}$, respectively.
Two approaches are considered to select the SOH levels of knots.
One is uniform knots where the SOHs up to EOL are uniformly divided, and the other is optimized knots, where Bayesian optimization~\cite{snoek2012practical} is used to find the optimal locations of the knots that can best construct the capacity trajectory.
Uniform knots are simple, intuitive, and easy to implement.
In contrast, optimized knots using Bayesian optimization may increase the accuracy of trajectory reconstruction.
\red{
A detailed explanation of Bayesian optimization is given in \ref{appendix:bo}.}

%
For automatic feature selection, we propose the sequential data processing method.
First, raw sequential data were obtained for the voltage, current, and corresponding time during the charging and discharging processes of the battery.
Because the voltage and current are measured at regular time intervals during the battery charging and discharging process, the data dimensions may vary widely depending on the difference in the charging and discharging times for each battery and cycle number.
Therefore, to match the dimensions of the input data applied to the CNN model, each variable of the charging data was uniformly discretized into $N$ equal parts over time.
Let the voltage $\mathbf{v}=[v_1, v_2, \ldots, v_L]^T$ and current $\mathbf{i}=[i_1, i_2, \ldots, i_L]^T$ be measured at time $\mathbf{t}=[t_1, t_2, \ldots, t_L]^T$ in one charge-discharge cycle, where $L$ is the number of measured points.
Initially, the time $\mathbf{t}$ is uniformly discretized into $N$ points, that is, $\tilde{\mathbf{t}}=[\tilde{t}_1, \tilde{t}_2, \ldots, \tilde{t}_N]^T$, where $\tilde{t}_n = t_1 + (t_L - t_1)\cdot (n-1)/(N-1)$, $n=1,2,\ldots,N$.
On $\tilde{\mathbf{t}}$, the voltage $\tilde{\mathbf{v}}=[\tilde{v}_1, \tilde{v}_2, \ldots, \tilde{v}_N]^T$ and current $\tilde{\mathbf{i}}=[\tilde{i}_1, \tilde{i}_2, \ldots, \tilde{i}_N]^T$ are obtained through linear interpolation.
Finally, the input to the model is processed into a matrix $\mathbf{X}=\left[\tilde{\mathbf{v}}~\tilde{\mathbf{i}}~\tilde{\mathbf{t}}\right]$.
The processed inputs implicitly represent the battery charge-discharge process for a certain period.
Creating a model input using the data processing method helps simplify the calculation process and has the advantage that it can be applied to batteries under various charging conditions.
In addition, the proposed pre-processing method helps prevent the first or last part of the voltage or current from being cut off.


\subsection{Model structure}\label{sec2.2:modelstructure}

%
%
To extract features from processed inputs, a CNN was adopted to predict the capacity trajectory of the batteries.
The key element that differentiates the CNN from standard neural networks or recurrent neural networks is the convolution layers.
The convolution layer extracts information from the data by calculating the convolution, where the kernel moves over the sequential data.
In addition, a fully connected (FC) layer is applied to the final part of the model, wherein the model output is mapped to values of interest.
In contrast to the manual feature selection, the CNN automatically learns the optimal features from raw input data.

%
%
In the capacity degradation trajectory prediction, the model predicts the locations of the knots that are used to construct the capacity trajectory, as illustrated in Fig.~\ref{fig:libschema}.
It is worth noting that the cycle numbers of knots for the given SOH levels are predicted instead of the capacity values of knots.
In addition, the $K$ knots always include the \red{EOL} point for robust construction of the capacity trajectory.
%
%
Letting the CNN model for capacity trajectory prediction be denoted as $f_{tr}$, the FC layer outputs $K$-dimensional real values $\mathbf{z}=[z_1, z_2, \ldots, z_K]^T$, and the exponential operation is applied to obtain positive values, which represents intervals $\mathbf{h}=[h_1, h_2, \ldots, h_K]^T$ in cycle numbers of knots in specific SOH levels, as shown in Fig.~\ref{fig:libschema}.
\begin{align}
    f_{tr}(\mathbf{X}) = \begin{bmatrix}
        \exp(z_1)\\\exp(z_2)\\\vdots\\\exp(z_K)
    \end{bmatrix} = \begin{bmatrix}
        h_1\\h_2\\\vdots\\h_K
    \end{bmatrix}.\label{eq:traj_output}
\end{align}

\subsection{Data description}

\begin{table}[t!]
    \centering
    \caption{Summary of datasets.}
    \label{tab:datasets}
    \scalebox{1.0}{
    \begin{tabular}{lll}
         \toprule
         Datasets & Dataset 1~\cite{severson2019data,attia2020closed} & Dataset 2~\cite{zhu2022data}\\
         \midrule
         Nominal capacity & 1.1 Ah & 3.5 Ah\\
         Cycling temperature &30 \textdegree C & 25, 35, 45 \textdegree C \\
         Cell chemistry & LFP & NCA, NCM \\
         & Graphite & Graphite \\
         Number of cells & 169 & 82\\
         {Charging profile} & Multistage CC-CV &  CC-CV (0.25, 0.5, 1 C)\\
         {Discharging profile} & CC (4 C) & CC (1, 2, 4 C)\\
         \bottomrule
    \end{tabular}
    }
\end{table}

\begin{figure}[t]
    \centering
    \subfigure[]{
    \centering
    \includegraphics[width=0.45\textwidth]{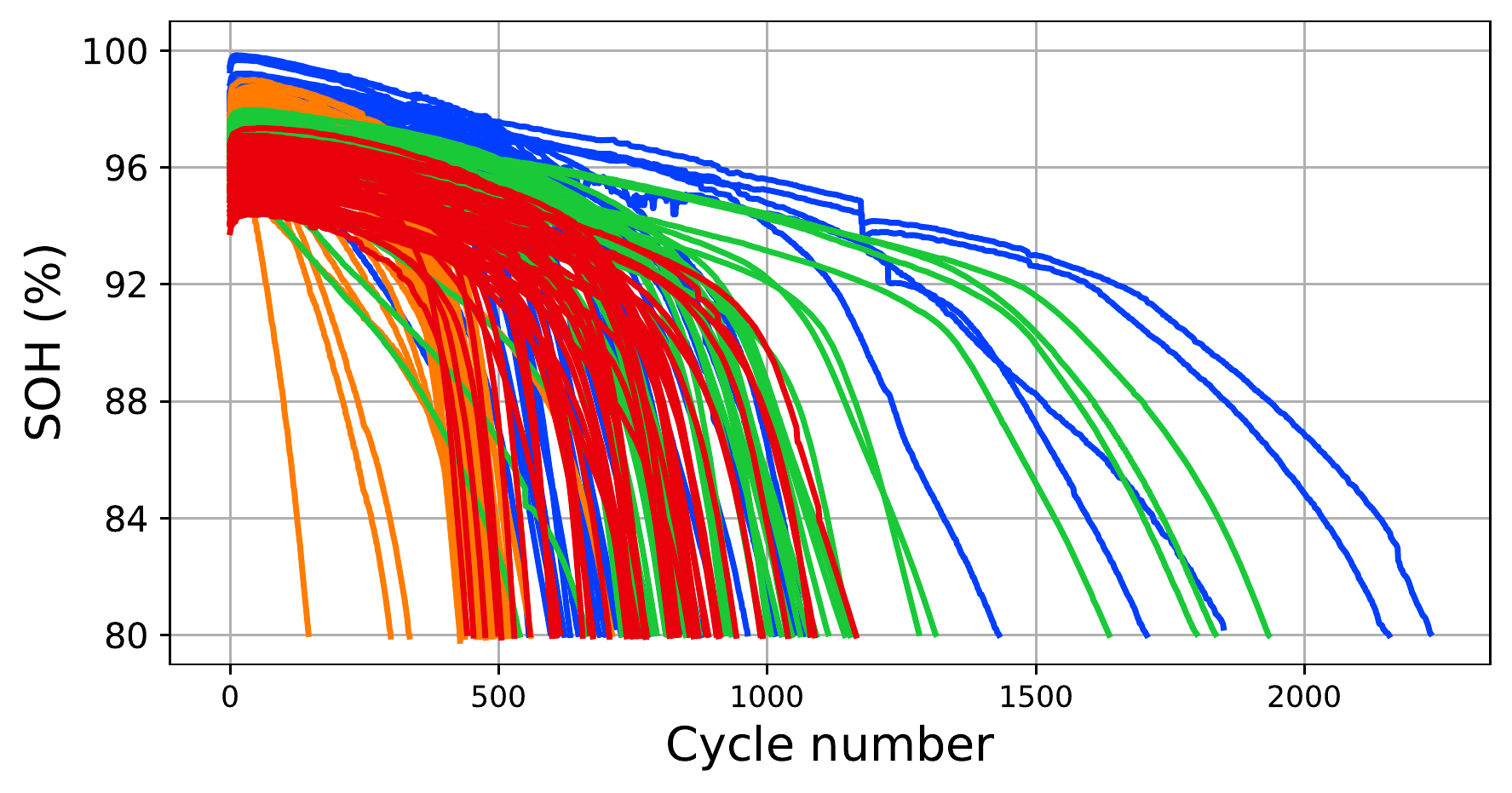}
    }
    \subfigure[]{
    \centering
    \includegraphics[width=0.45\textwidth]{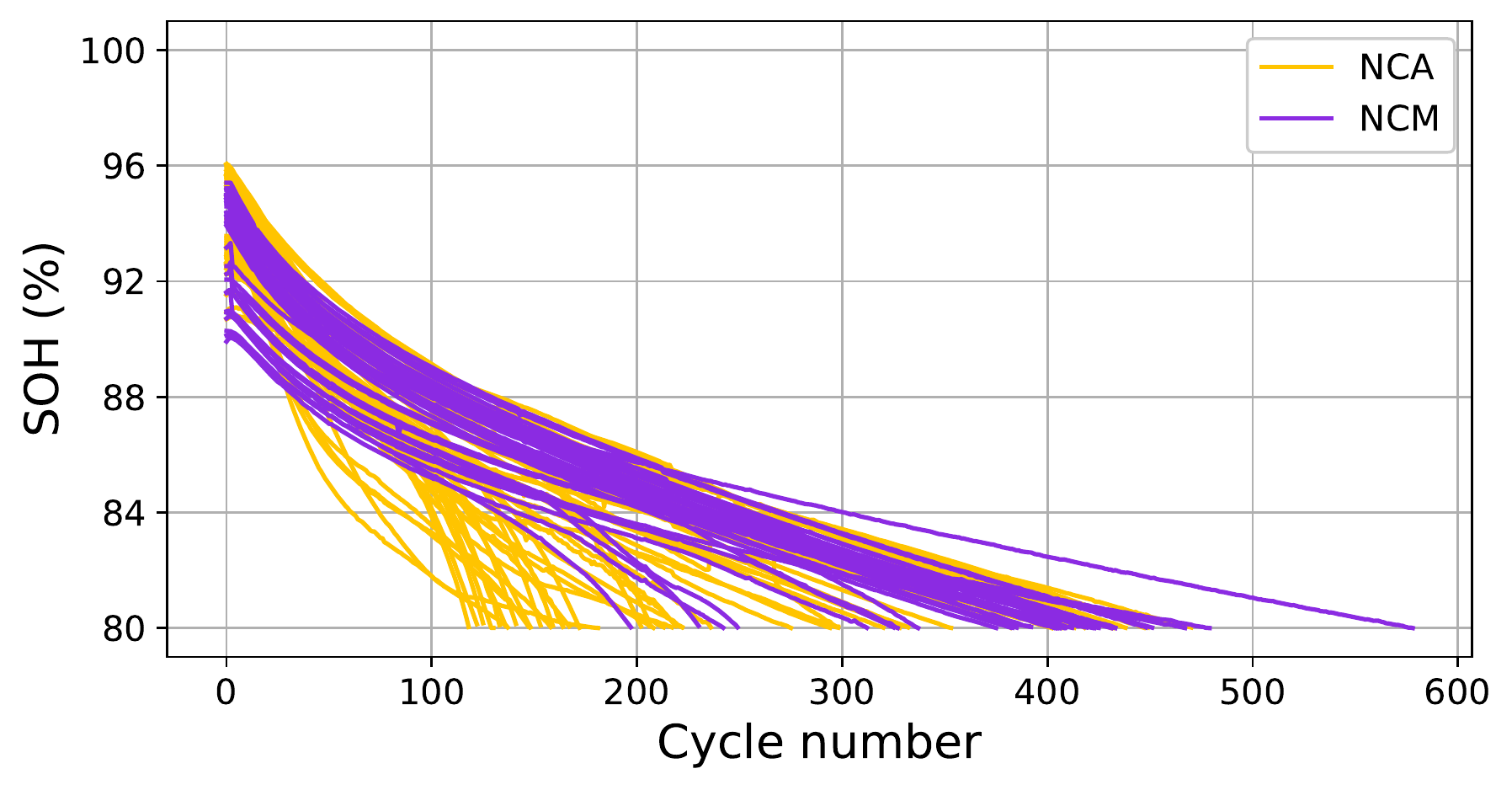}
    }
    \caption{Capacity degradation trajectory of experimental data from (a)~\cite{severson2019data,attia2020closed} and (b)~\cite{zhu2022data}. In (a), each color identifies a batch date, the date experiments started.}
    \label{fig:data_SOH}
\end{figure}

The proposed method was validated using two datasets from different sources, which are summarized in Table~\ref{tab:datasets}, and the capacity degradation trajectories of all the cells are shown in Fig.~\ref{fig:data_SOH}.
Dataset 1 involves cycling data under fast charging generated by Severson \textit{et al}.~\cite{severson2019data} and Attia \textit{et al}.~\cite{attia2020closed}, which contained 169 commercial lithium iron phosphoric (LFP) / graphite cells with a nominal capacity of 1.1 Ah.
All the cells were cycled in a temperature chamber set at 30 \textdegree C until EOL, which refers to 80\% of the nominal capacity.
The cell charging profiles include various forms of multistage constant current (CC).
The cells were charged with a one-, two-, or four-step fast-charging policy up to 80\% SOC.
After that, the cells were charged under a uniform constant current-constant voltage (CC-CV) of 1 C to a cutoff voltage of 3.6 V and cutoff current of C/50.
All cells were subsequently discharged at 4 C to 2.0 V.
Even though all batteries were manufactured on the same date, slight differences in the initial capacity occurred due to the calendar aging according to the experimental schedule, as shown in Fig.~\ref{fig:data_SOH}~(a).
The shapes of trajectories vary widely depending on the battery, while most batteries have a cycle life between 500 and 1,100 cycles.
Dataset 2 includes 130 commercial lithium-ion ion cells, containing 66 lithium nickel cobalt aluminum (NCA) cells, 55 lithium nickel cobalt manganese (NCM) cells, and nine NCM$+$NCA cells cycled under various conditions, generated by Zhu \textit{et al}.~\cite{zhu2022data}.
The nominal capacities of batteries are 3.5 Ah for NCA and NCM and 2.5 Ah for NCM$+$NCA.
All the cells were cycled in a temperature chamber of 25, 35, and 45 \textdegree C.
The cell cycling was performed with CC charging to 4.2 V with various current rates from 0.25 C to 1 C, and CV charging was followed at 4.2 V until a current of 0.05 C was reached.
After 30 minutes of relaxation, CC was employed for discharging to 2.65 V for the NCA cells and 2.5 V for the NCM cells.
Note that the cells that did not reach 80\% or with a cycle number less than 30 were excluded from the dataset, and the remaining 82 cells of NCA and NCM were used for validation.
The cycle number of the batteries is between 100 and 600, and the NCA and NCM batteries show different shapes of the trajectory, as shown in Fig.~\ref{fig:data_SOH}~(b).
In addition, the degradation trends of the batteries are dissimilar from those in Dataset 1.
It is worth noting that Dataset 1 was used to verify the effectiveness of the proposed method and for a detailed comparison with existing methods, and Dataset 2 was used to demonstrate the applicability of the proposed method to degradation curves of various patterns.
\red{
In this study, the SOH is defined as the ratio of discharge capacity to the nominal capacity, where discharge capacity can be calculated by integrating discharge current with time.}

\subsection{Implementation details}

The proposed CNN comprises a convolution block and an FC layer.
A convolution block includes a convolution layer, batch normalization, rectified linear unit (ReLU) activation function, and dropout layer in a single set.
The Ranger optimizer~\cite{wright2021ranger21}, which combines the rectified adaptive moment estimation (RAdam) optimizer~\cite{liu2019variance} with the look-ahead method~\cite{zhang2019lookahead}, was used for model training.
The loss function was set to cross-entropy for classification problems and the mean absolute error (MAE) for regression problems.
We then validated all experimental data using five-fold cross-validation, where the method was first implemented by dividing the entire data into five classes equally for the size of the knee point and the experiment dates, followed by a total of five iterations by dividing the training and test sets with a 4:1 ratio.
The five-fold cross-validation is a widely-adopted validation method for machine learning and neural networks that ensures that all data are used for testing, thus avoiding the class imbalance issue and maximizing the data used in training~\cite{hsu2022deep, shen2020deep}.
It is worth noting that the input variables and hyperparameters of the network were carefully chosen with a case study, which is described in \ref{appendix:tuning}.
The voltage, current, and time were selected as the input variables.
The kernel size of the convolution layer, batch size, and learning rate were set as 4, 32, and 0.001, respectively.
Four convolution blocks with output channel sizes 4, 8, 16, and 32 were stacked, and the outputs of the convolution blocks were flattened and fed to the FC layer.
\red{
The proposed architecture of the CNN model is described in Table~\ref{tab:CNN} and} was implemented using PyTorch~\cite{paszke2019pytorch}.
\red{
Furthermore, Monte Carlo (MC) dropout~\cite{gal2015bayesian} was adopted to estimate the predictive distribution to obtain the mean and confidence interval (CI).
The detailed implementation of MC dropout on a CNN is explained in Gal and Ghahramani~\cite{gal2015bayesian}.}
The procedure for the proposed capacity trajectory prediction method is summarized as follows:

\begin{enumerate}
    \item Data pre-processing and model training:

    \begin{enumerate}
        \item Data pre-processing: A sequential data processing method was applied to raw data for the voltage, current, and corresponding time, which is described in Section~\ref{sec2.1:preprocessing}.
        The raw sequential battery data during the charging and discharging processes in a single cycle were uniformly discretized into $N=128$ intervals.

        \item Knot identification: Through uniform division or Bayesian optimization, the SOH levels are obtained using training data.
        It is worth noting that the EOL is always included in the knots.
        %

        \item Model training: The CNN model is constructed with the processed data as the input and the interval of knots as the output.
        In particular, the output dimension of the last FC layer of the CNN model was equal to the number of knots.
        The loss function was set to the MAE between the predicted and observed cycle numbers of the knots.

    \end{enumerate}

    \item Model validation:

    \begin{enumerate}
        \item Knot prediction: Given the raw data of the first single cycle of a test cell, the data were uniformly discretized and applied to the trained CNN model.
        The intervals in the cycle numbers of the knots are then obtained, as shown in \eqref{eq:traj_output}.
        The cycle numbers of knots $\mathbf{p} = [p_1, p_2, \ldots, p_K]^T$ are calculated by summing the intervals $\mathbf{h}$ cumulatively, as follows:
        \begin{align}
            p_k = \sum_{j=1}^k h_j,~k=1,2,\ldots,K.\label{eq:cumulativeknots}
        \end{align}

        \item Capacity trajectory construction: The capacity trajectory was constructed based on the predicted knots using the PCHIP interpolation.
        PCHIP interpolation was implemented using \texttt{SciPy}~\cite{2020SciPy-NMeth}.

    \end{enumerate}

\end{enumerate}

To evaluate the performance of the proposed method, the MAE and mean absolute percentage error (MAPE) of the prediction were calculated as follows.
\begin{align}
    MAE&=\frac{1}{n}\sum_{i=1}^{n} \left| y_i - \hat{y_i} \right|.\label{eq:MAE}\\
    MAPE&=\frac{1}{n}\sum_{i=1}^{n} \frac{ \left| y_i - \hat{y_i} \right|}{y_i}.\label{eq:MAPE}
\end{align}
where $y_i$ and $\hat y_{i}$ denote the observed and predicted measures (i.e., cycle number and capacity) at the $i$th index, respectively.
%
%

\begin{table}[t!]
    \centering
    \caption{\red{Proposed architecture of the CNN model. Hyperparameters represent kernel size, stride, and padding for convolution layers; activation function for activation layers; and dropout rate for dropout layers.}}
    \label{tab:CNN}
    \scalebox{1.0}{
    \begin{tabular}{llll}
         \toprule
         Layer & Input size & Hyperparameters & Output size\\
         \midrule
         Convolution 1 & $3\times 128$ & 4, 2, 1 & $4\times 64$\\
         Batch normalization 1 & $4\times 64$ & & $4\times 64$\\
         Activation 1 & $4\times 64$ & ReLU & $4\times 64$\\
         Dropout 1 & $4\times 64$ & 0.2 & $4\times 64$\\
         Convolution 2 & $4\times 64$ & 4, 2, 1 & $8\times 32$\\
         Batch normalization 2  & $4\times 64$ &  & $8\times 32$\\
         Activation 2  & $4\times 64$ & ReLU & $8\times 32$\\
         Dropout 2  & $4\times 64$ & 0.2 & $8\times 32$\\
         Convolution 3 & $8\times 32$ & 4, 2, 1 & $16\times 16$\\
         Batch normalization 3 & $8\times 32$ & & $16\times 16$\\
         Activation 3 & $8\times 32$ & ReLU & $16\times 16$\\
         Dropout 3 & $8\times 32$ & 0.2 & $16\times 16$\\
         Convolution 4 & $16\times 16$ & 4, 2, 1 & $32\times 8$\\
         Batch normalization 4 & $16\times 16$ &  & $32\times 8$\\
         Activation 4 & $16\times 16$ & ReLU & $32\times 8$\\
         Dropout 4 & $16\times 16$ & 0.2 & $32\times 8$\\
         Flatten & $32\times 8$ & & 256\\
         Linear & 256 & & $K$ \\
         Activation 5 & $K$ & Exponential & $K$\\
         \bottomrule
    \end{tabular}
    }
\end{table}


\section{Results and discussion}\label{sec3:results}

\subsection{\red{Knot-based trajectory prediction}}

The SOH levels of the knots were first determined for Dataset 1 in order to predict the capacity degradation trajectory~\cite{severson2019data, attia2020closed}.
It is worth recalling that the uniform knots are simply $K$ divisions of the SOH level up to EOL (80\% SOH), and the optimized knots were obtained using Bayesian optimization by minimizing the interpolation error $d(\mathbf{p})$ between the constructed and observed trajectories \red{of the training cells}, as shown in \eqref{eq:BO_loss}.
The SOH levels of uniform and optimized knots were obtained in accordance with the number of knots $K=2,3$, and $4$, as listed in Table~\ref{tab:SOHcurve_optim} with interpolation error $d(\mathbf{p})$.
\red{
Note that locations for optimized knots are listed for every fold, because the training and test sets can vary during five-fold cross-validation, and only $d(\mathbf{p})$ for the test set is listed.}
In addition, all the knots included a predefined SOH level of EOL, that is, 80\%.
The interpolation error in the optimized knots was less than that in the uniform knots for $K=2$ and $K=3$ and greater for $K=4$, but the difference was insignificant.
This indicates that a smaller error is seen when constructing the capacity trajectory by obtaining knots using Bayesian optimization than when using uniform knots.
Meanwhile, the interpolation error decreases in both the uniform and optimized knots as the number of knots increases owing to an increased degree of freedom.

\begin{table}[t]
\centering
\caption{SOH levels of uniform and optimized knots with interpolation error $d(\mathbf{p})$ for $K=2,3,4$. \red{Note that locations for optimized knots are listed for every fold, because the training and test sets can vary during five-fold cross-validation, and only $d(\mathbf{p})$ for the test set is listed.}}
\begin{adjustbox}{width=0.95\textwidth}
\begin{tabular}{lccccccc}
\toprule
& Number of knots $K$ & Fold & \multicolumn{4}{c}{SOH levels for knots (\%)} & Interpolation error $d(\mathbf{p})$\\
\midrule
Uniform knots & 2 & - & 80.0 & 89.0 &  && $8.50\times10^{-3}$\\
& 3 & - & 80.0 & 86.0 & 92.0 && $4.65\times10^{-3}$\\
& 4 & - & 80.0 & 84.5 & 89.0 & 93.5 & $3.39\times10^{-3}$ \\
\midrule
Optimized knots & 2 & 1  & 80.0 & 91.9 &  && $7.16\times10^{-3}$\\
&& 2  & 80.0 & 91.4 &  && $5.93\times10^{-3}$\\
&& 3  & 80.0 & 90.6 &  && $6.76\times10^{-3}$\\
&& 4  & 80.0 & 90.6 &  && $6.51\times10^{-3}$\\
&& 5  & 80.0 & 90.5 &  && $6.55\times10^{-3}$\\
& 3 & 1 & 80.0 & 90.4 & 93.8 && $3.44\times10^{-3}$\\
&& 2 & 80.0 & 89.1 & 93.8 && $3.44\times10^{-3}$\\
&& 3 & 80.0 & 89.1 & 93.8 && $3.36\times10^{-3}$\\
&& 4 & 80.0 & 89.1 & 93.8 && $3.38\times10^{-3}$\\
&& 5 & 80.0 & 89.1 & 93.8 && $3.27\times10^{-3}$\\
& 4 & 1 & 80.0 & 83.7 & 89.8 & 94.0 & $3.00\times10^{-3}$ \\
&& 2 & 80.0 & 83.7 & 89.8 & 94.0 & $3.05\times10^{-3}$ \\
&& 3 & 80.0 & 83.7 & 89.8 & 94.0 & $2.96\times10^{-3}$ \\
&& 4 & 80.0 & 86.2 & 91.4 & 95.0 & $2.55\times10^{-3}$ \\
&& 5 & 80.0 & 83.7 & 89.8 & 94.0 & $2.89\times10^{-3}$ \\
\bottomrule
\end{tabular}
\end{adjustbox}
\label{tab:SOHcurve_optim}
\end{table}

\begin{figure}[t]
    \centering
    \subfigure[]{
    \centering
    \includegraphics[width=0.31\textwidth]{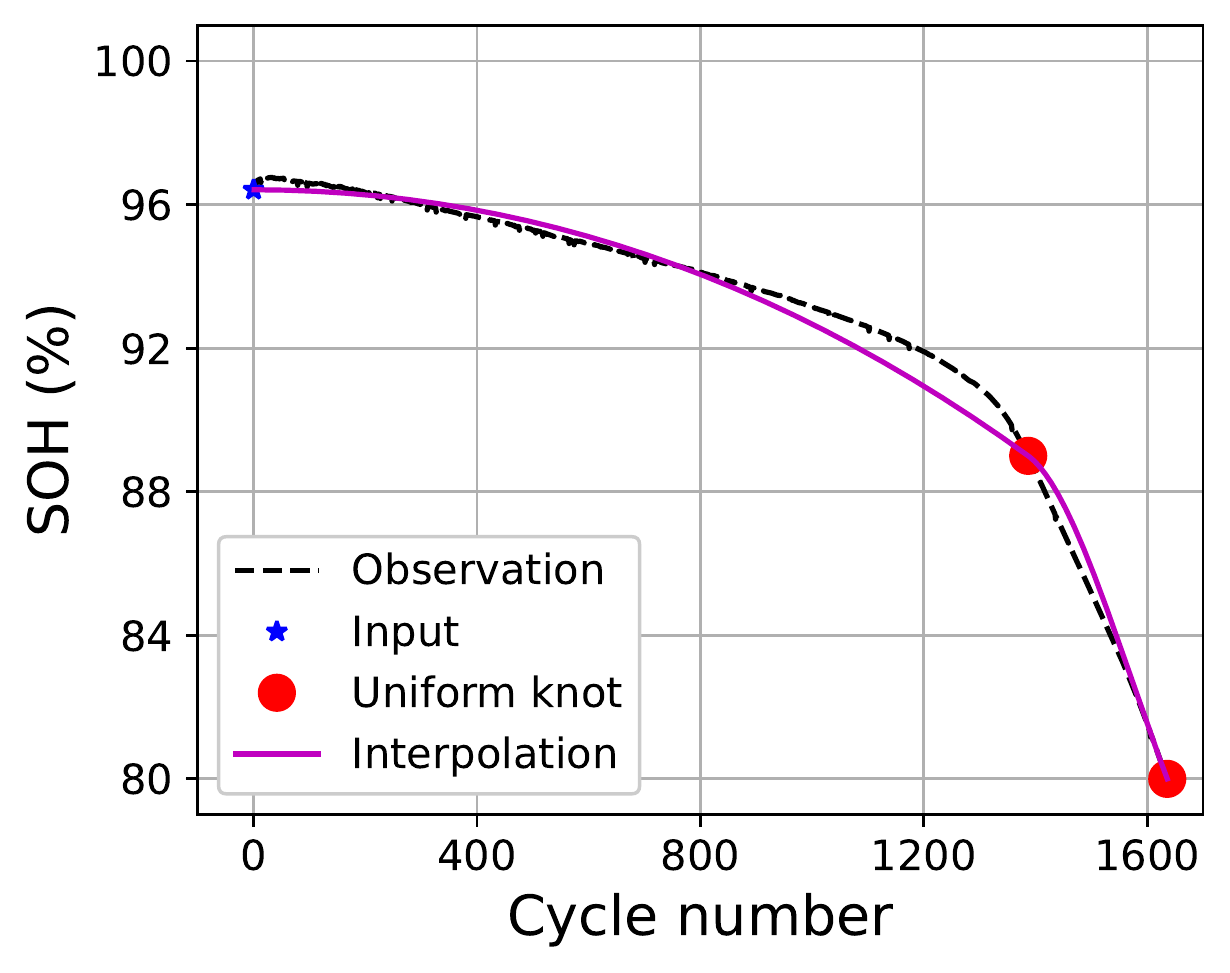}
    }
    \subfigure[]{
    \centering
    \includegraphics[width=0.31\textwidth]{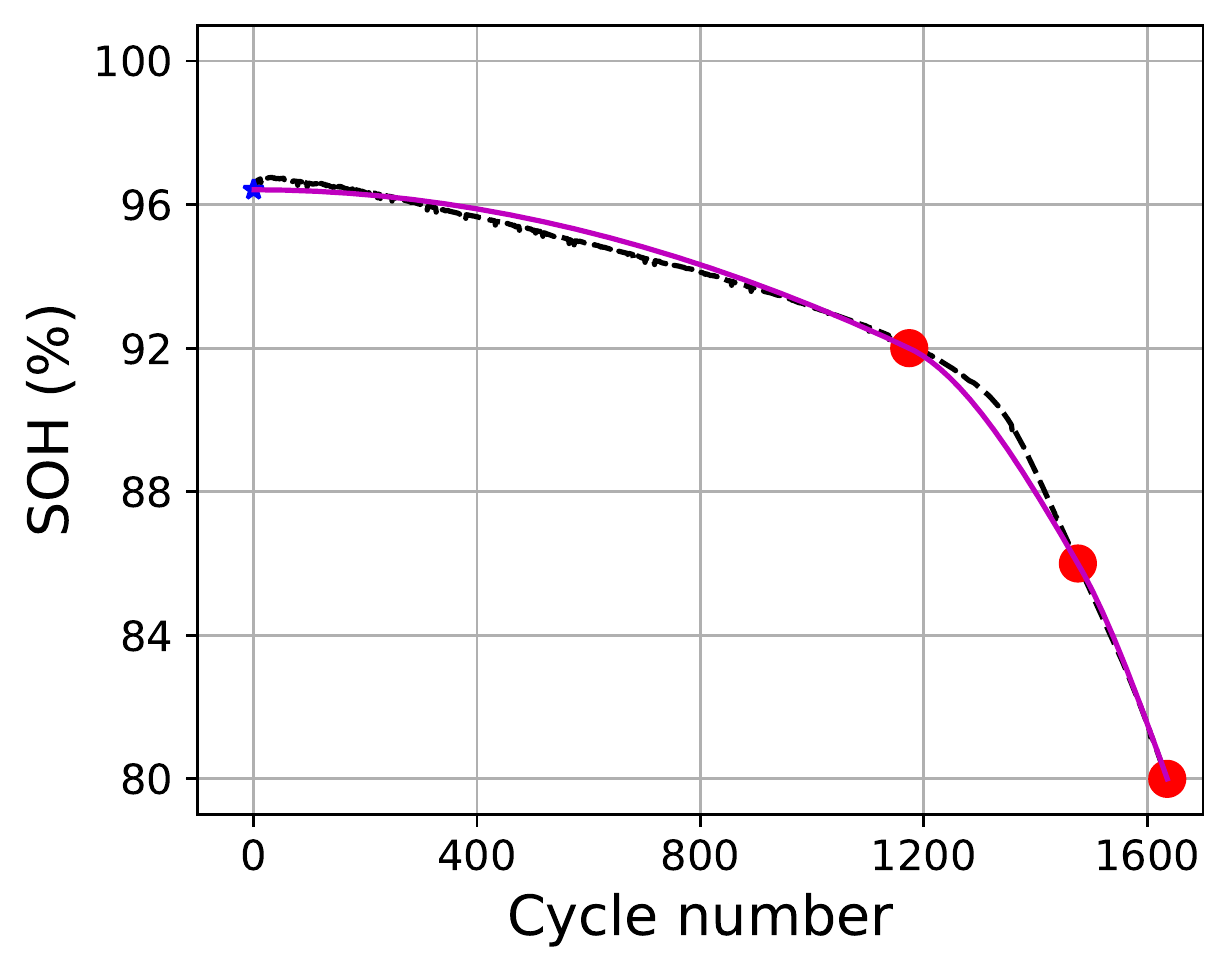}
    }
    \subfigure[]{
    \centering
    \includegraphics[width=0.31\textwidth]{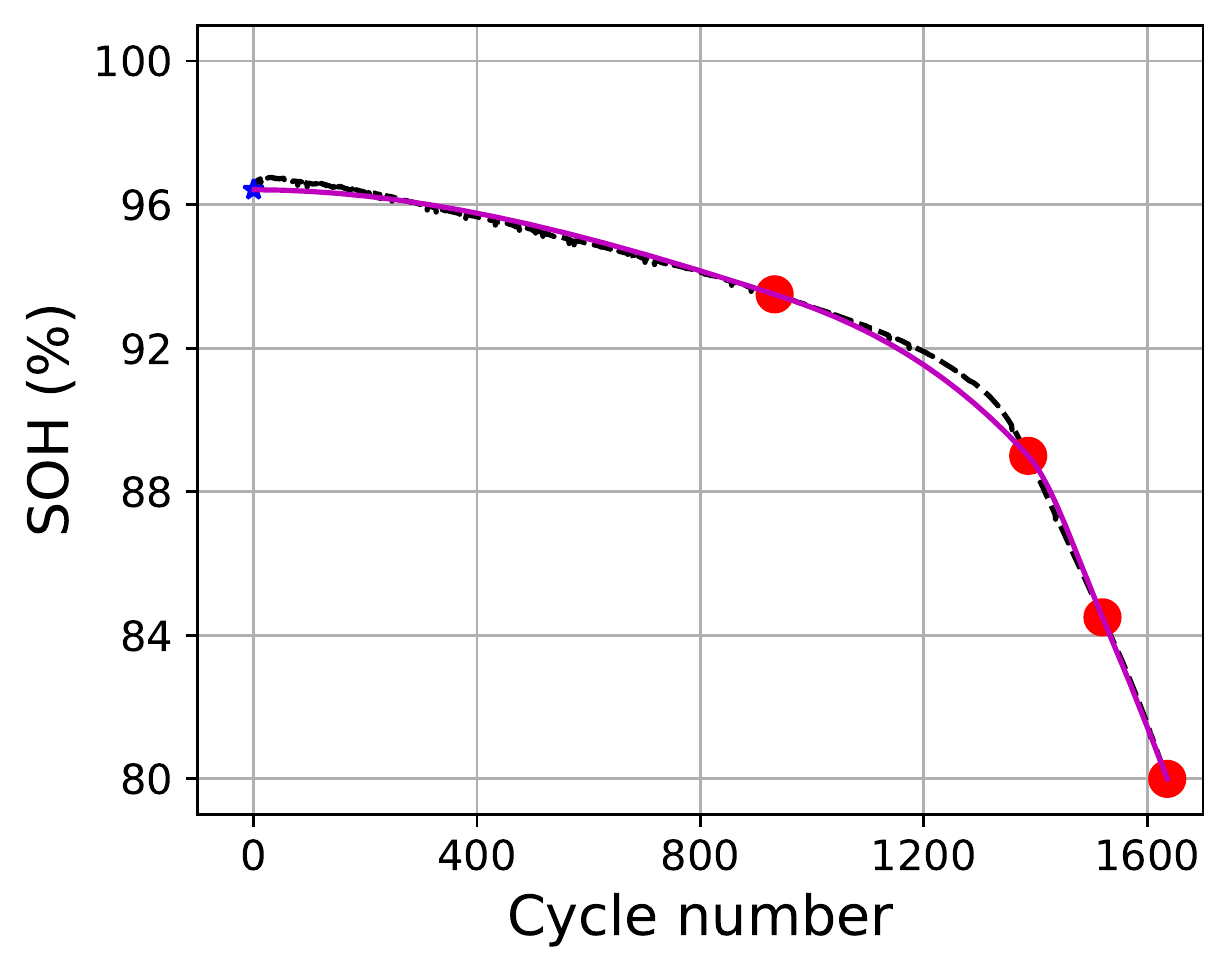}
    }
    \subfigure[]{
    \centering
    \includegraphics[width=0.31\textwidth]{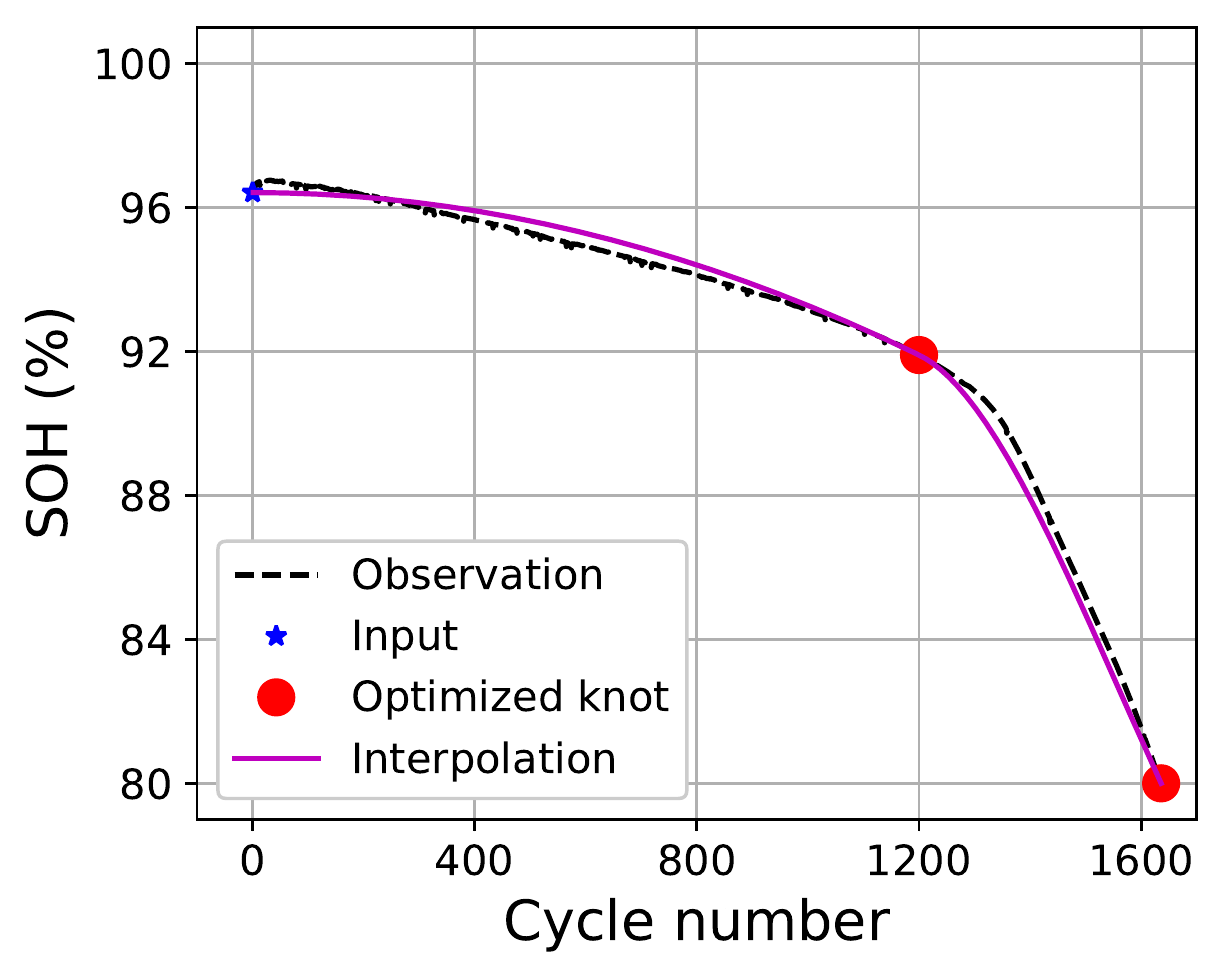}
    }
    \subfigure[]{
    \centering
    \includegraphics[width=0.31\textwidth]{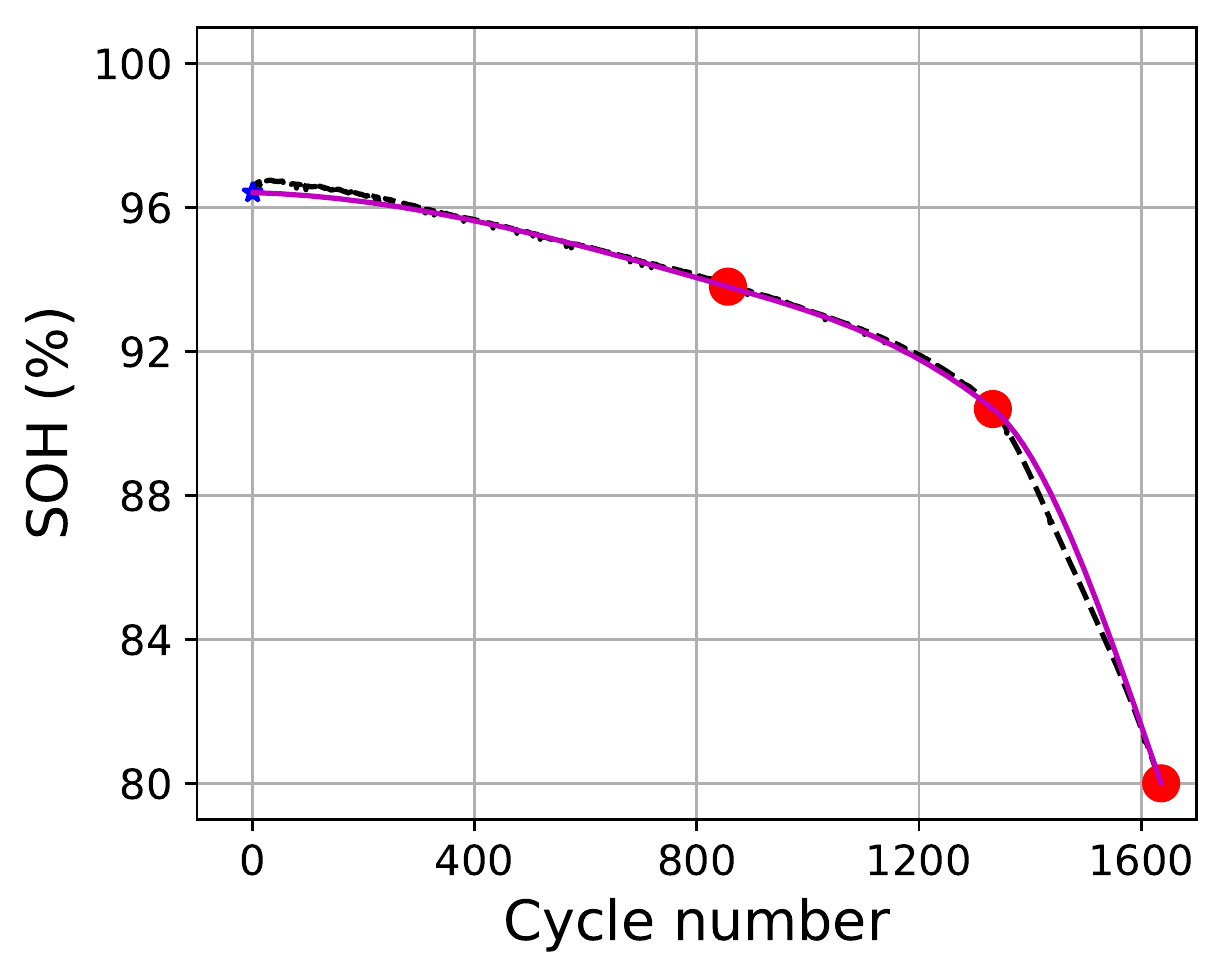}
    }
    \subfigure[]{
    \centering
    \includegraphics[width=0.31\textwidth]{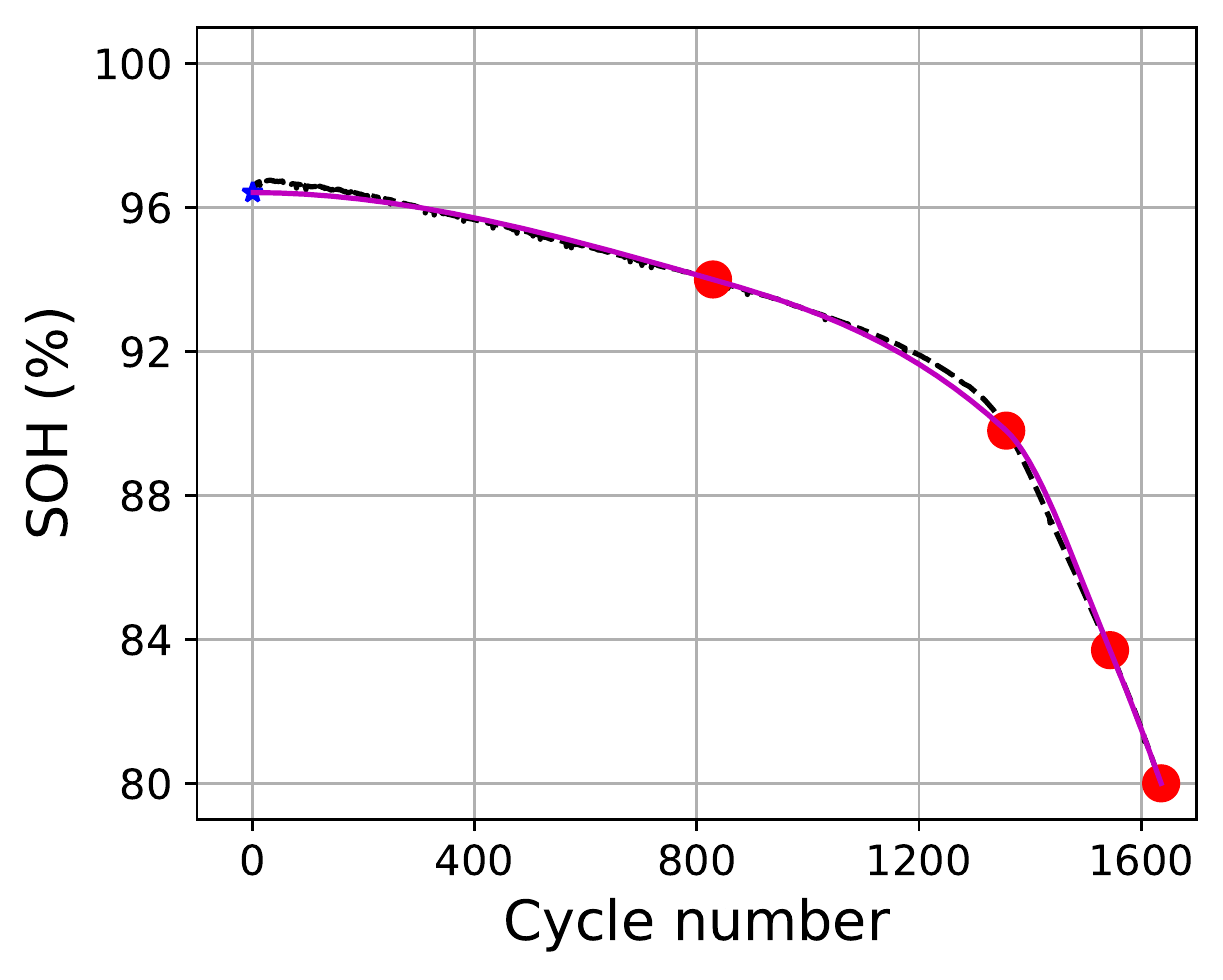}
    }
    \caption{Examples of (a--c) uniform and (d--f) optimized knots with reconstructed capacity trajectories for $K=2,3,4$.}
    \label{fig:SOH_optim}
\end{figure}

A few examples of the reconstructed capacity trajectories using PCHIP interpolation on uniform and optimized knots are illustrated in Fig.~\ref{fig:SOH_optim}.
The reconstructed capacity trajectories on the uniform knots are shown in Fig.~\ref{fig:SOH_optim}~(a)--(c) for $K=2,3,$ and $4$, respectively.
Because the knots are uniformly divided according to the SOHs, they are located where the capacity drops sharply in the case of a concave capacity trajectory.
%
A few examples of the reconstructed capacity trajectories on the optimized knots \red{in fold 1} are illustrated in Fig.~\ref{fig:SOH_optim}~(d)--(f).
Unlike uniform knots, optimized knots are located in the rapidly decreasing region of the capacity trajectory for all the numbers of knots.
These results imply that knee points and knee onsets are primary indicators of the capacity trajectory.
Furthermore, for $K=2$, the error was larger before than after the knot of the capacity trajectory for a uniform knot and smaller for an optimized knot.
This is because the knot location differs depending on whether it is located before or after the sharp capacity drop.
However, as $K$ increases, the interpolation error gradually decreases.
Therefore, when $K = 4$, the reconstructed trajectory almost perfectly mimics the observed trajectory in both uniform and optimized knots.
Thus, it was confirmed that the capacity trajectories can be constructed with a few knots.


The proposed CNN-based model receives the processed data of an early single cycle as input and outputs the cycle numbers of $K$ knots.
Subsequently, the entire capacity trajectory is constructed by interpolating the predicted knots.
\red{
Note that to evaluate the feature extraction performance of the CNN model, knee point classification and prediction using the CNN is validated through comparison with existing literature, which is described in \ref{sec:kneepoint_prediction}.}
The MAE and MAPE of the predicted knots in cycle numbers and the predicted trajectory in capacity using the proposed model on uniform and optimized knots are listed in Table~\ref{tab:knotprediction_results}.
It is confirmed that there are no significant differences in the MAE and MAPE according to $K$ for either knot prediction or trajectory prediction.
The left part of Table~\ref{tab:knotprediction_results} lists the results of the knot prediction.
For $K=2$, the MAEs (MAPEs) of the first- and second-predicted optimized knots were 70 (12.12\%) and 80 (9.82\%) cycles, respectively, while 70 (10.22\%) and 79 (9.53\%) cycles were required for uniform knots.
This result indicates that for both uniform and optimized knots, the first knot is relatively well predicted, but the MAPE is larger for the first knot because the observed cycle numbers are smaller than those for the second knot.
%
Similar to the case of $K=2$, the larger the knot cycle number, the larger the MAE, whereas the smaller the MAPE for $K=3$ and $4$.
An increase in MAE is assumed because of the output calculation in the proposed model, where the outputs of the model are cumulatively added, as shown in \eqref{eq:cumulativeknots}.
As knots with large observed cycle numbers are calculated by adding the preceding $h_k$, their prediction errors can accumulate.
The right portion of Table~\ref{tab:knotprediction_results} lists the MAE and MAPE for evaluating the interpolated capacity trajectory using the predicted knots.
Regardless of the number of knots $K$, the MAEs and MAPEs are less than 0.0140 Ah and 1.40\%, respectively, except for the uniform knots with $K=2$, which implies that it is difficult to express the trajectory with only two uniform knots.
In summary, the results indicate that the total capacity trajectory is predicted with high accuracy, even with only a few knots, implying that the two optimized knots are sufficient to predict the capacity trajectory.
Furthermore, it is confirmed that there is no significant difference between uniform and optimized knots, so the use of uniform knots is recommended owing to the simple and intuitive selection of SOH levels.

\begin{table}[t]
\centering
\caption{MAE and MAPE in optimized knot prediction and capacity trajectory prediction using the proposed model for $K=2,3,4$.}
\begin{adjustbox}{width=0.95\textwidth}
\begin{tabular}{lccccc}
\toprule
\multirow{2}{*}{} & \multirow{2}{*}{Number of knots $K$} & \multicolumn{2}{c}{Knot prediction} & \multicolumn{2}{c}{Trajectory prediction}\\
\cmidrule{3-6}
&& MAE (Cycle) & MAPE (\%) & MAE (Ah) & MAPE (\%)\\
\midrule
Uniform knots & 2   & 75 & 9.88 & 0.0153 & 1.51 \\
&3 & 78 &  11.54 & 0.0135 & 1.35 \\
&4  & 83 & 13.02 & 0.0135 & 1.35 \\
\midrule
Optimized knots & 2   & 81 & 11.21 & 0.0129 & 1.41 \\
&3 & 85 &  13.76 & 0.0128 & 1.40 \\
&4 & 97 & 16.47 & 0.0143 & 1.57 \\
\bottomrule
\end{tabular}
\label{tab:knotprediction_results}
\end{adjustbox}
\end{table}

\begin{figure}[t]
    \centering
    \subfigure[]{
    \centering
    \includegraphics[width=0.31\textwidth]{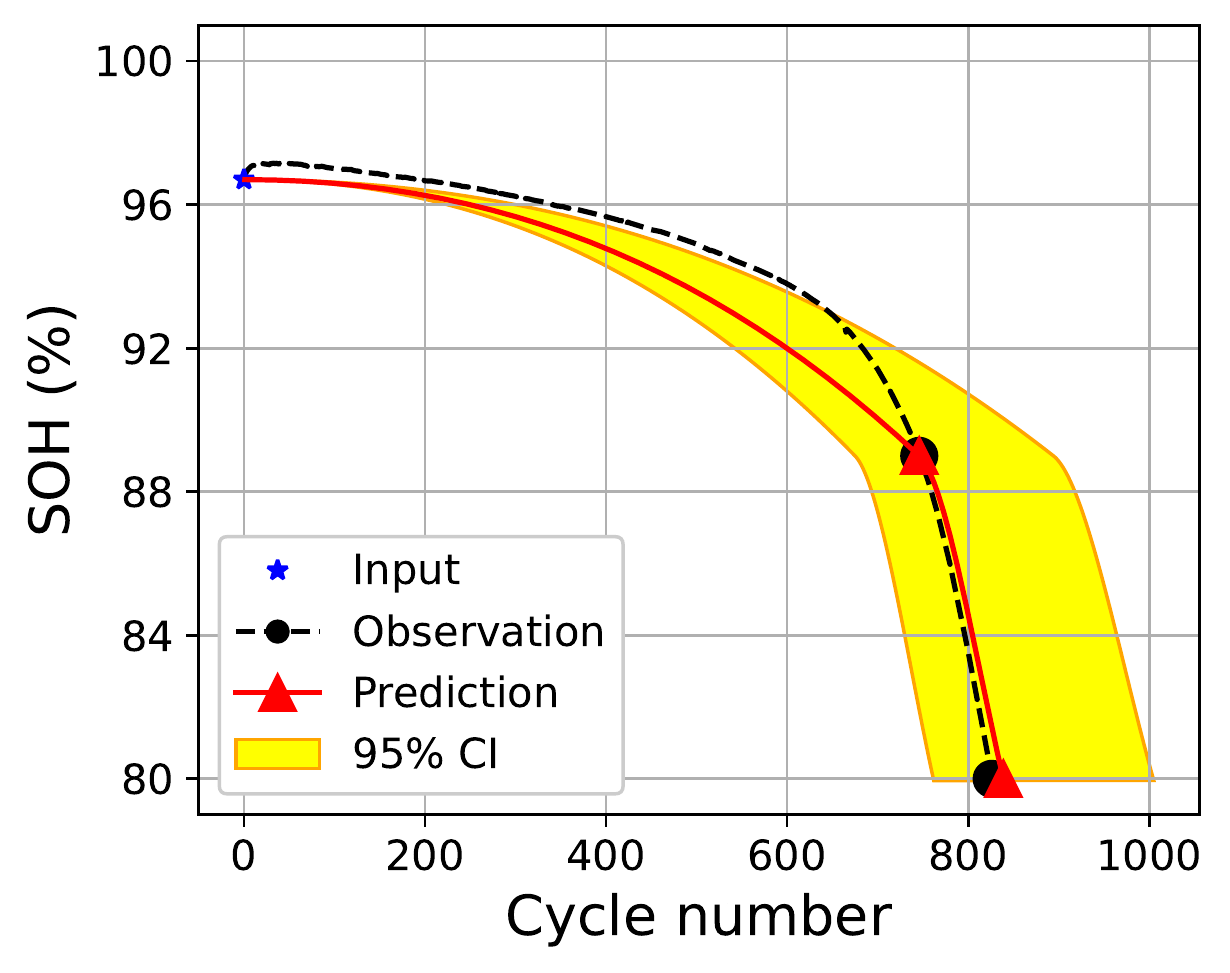}
    }
    \subfigure[]{
    \centering
    \includegraphics[width=0.31\textwidth]{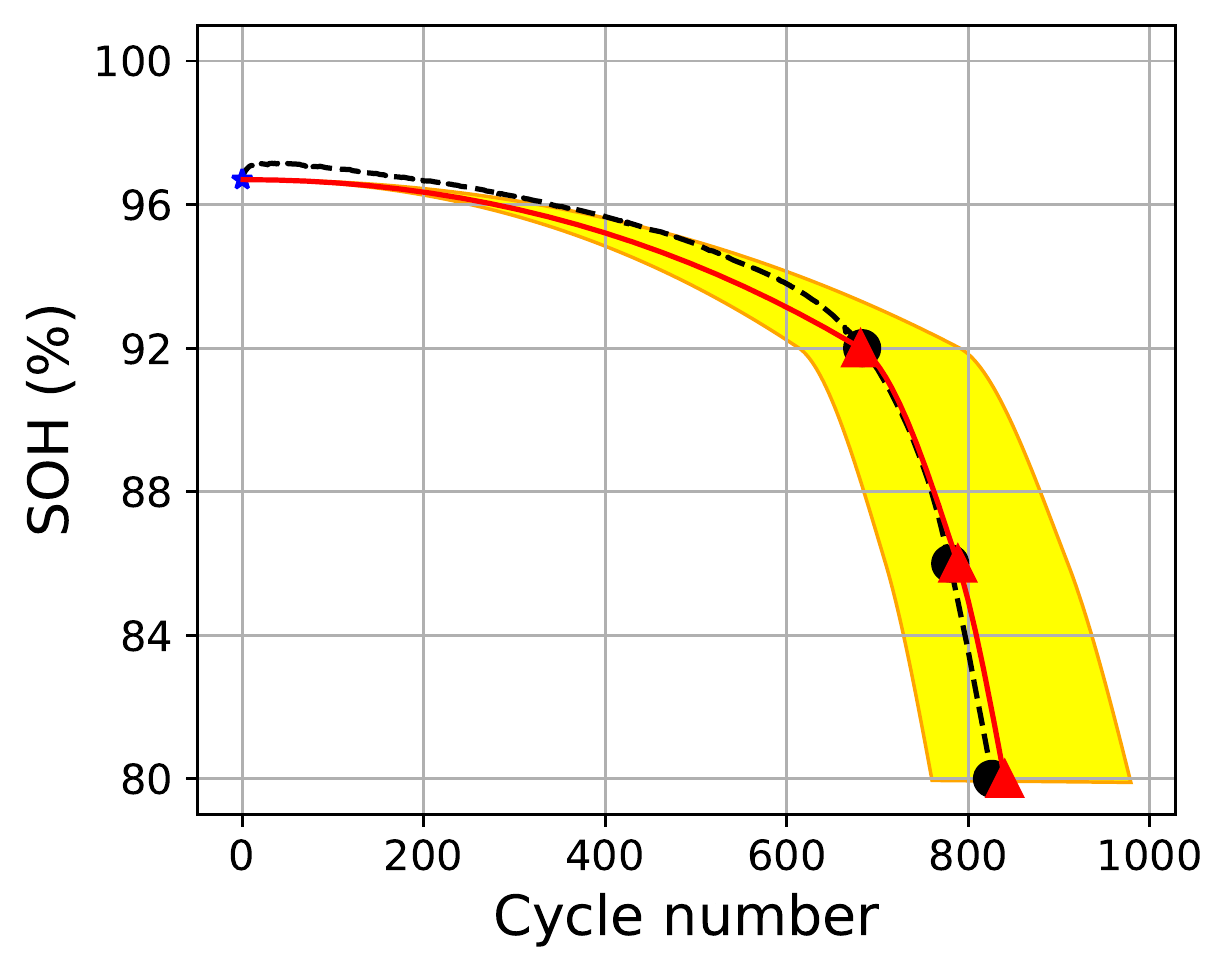}
    }
    \subfigure[]{
    \centering
    \includegraphics[width=0.31\textwidth]{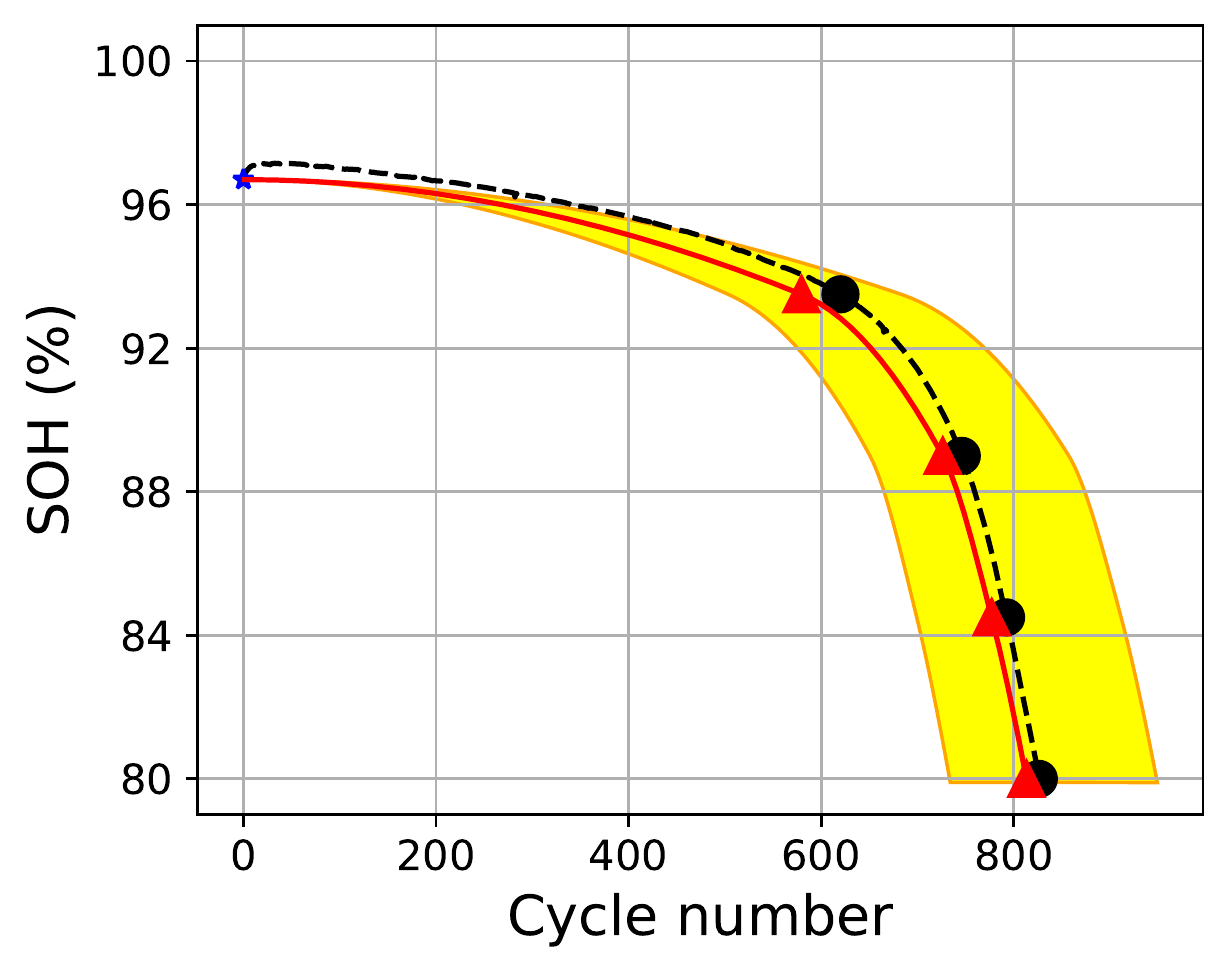}
    }
    \subfigure[]{
    \centering
    \includegraphics[width=0.31\textwidth]{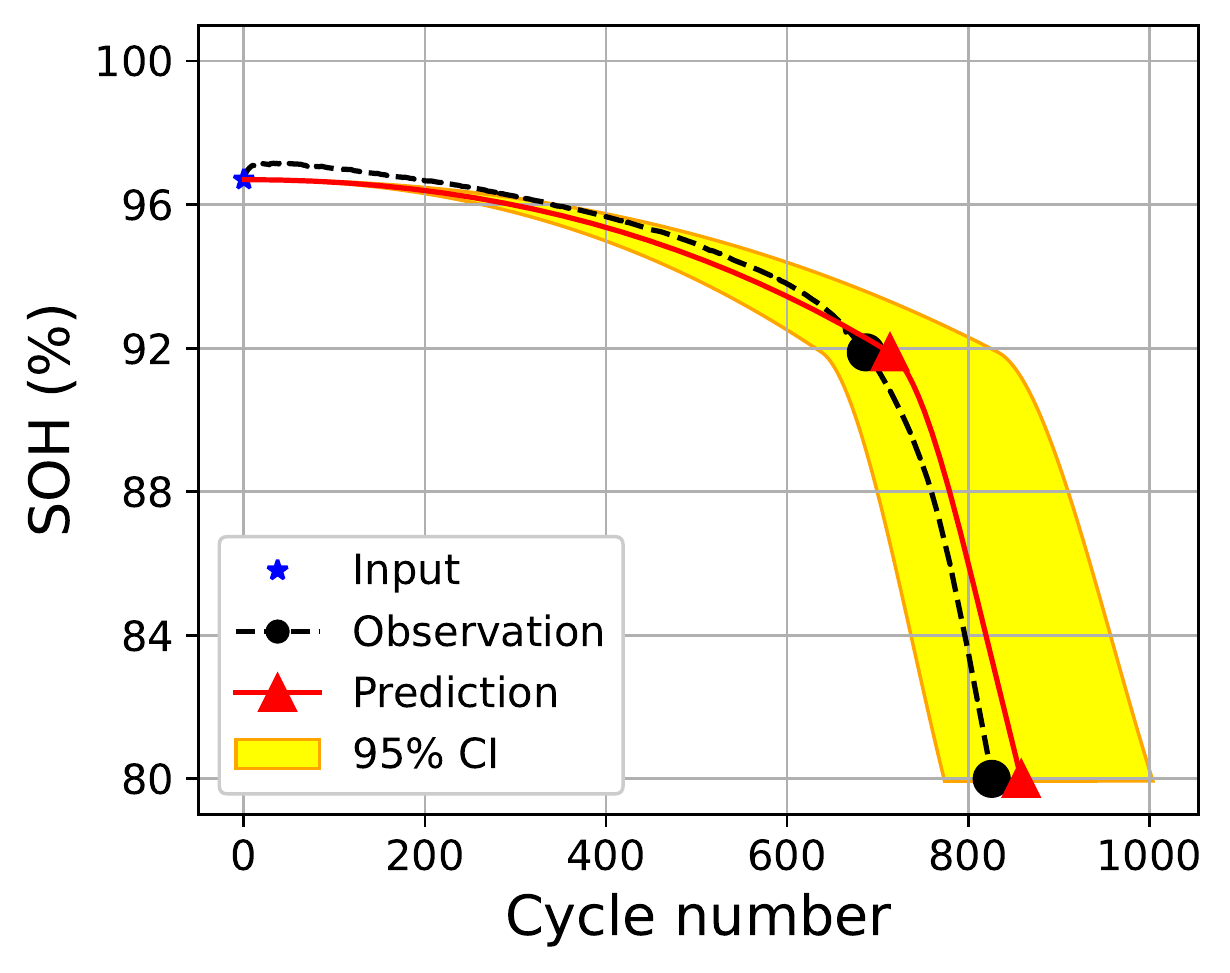}
    }
    \subfigure[]{
    \centering
    \includegraphics[width=0.31\textwidth]{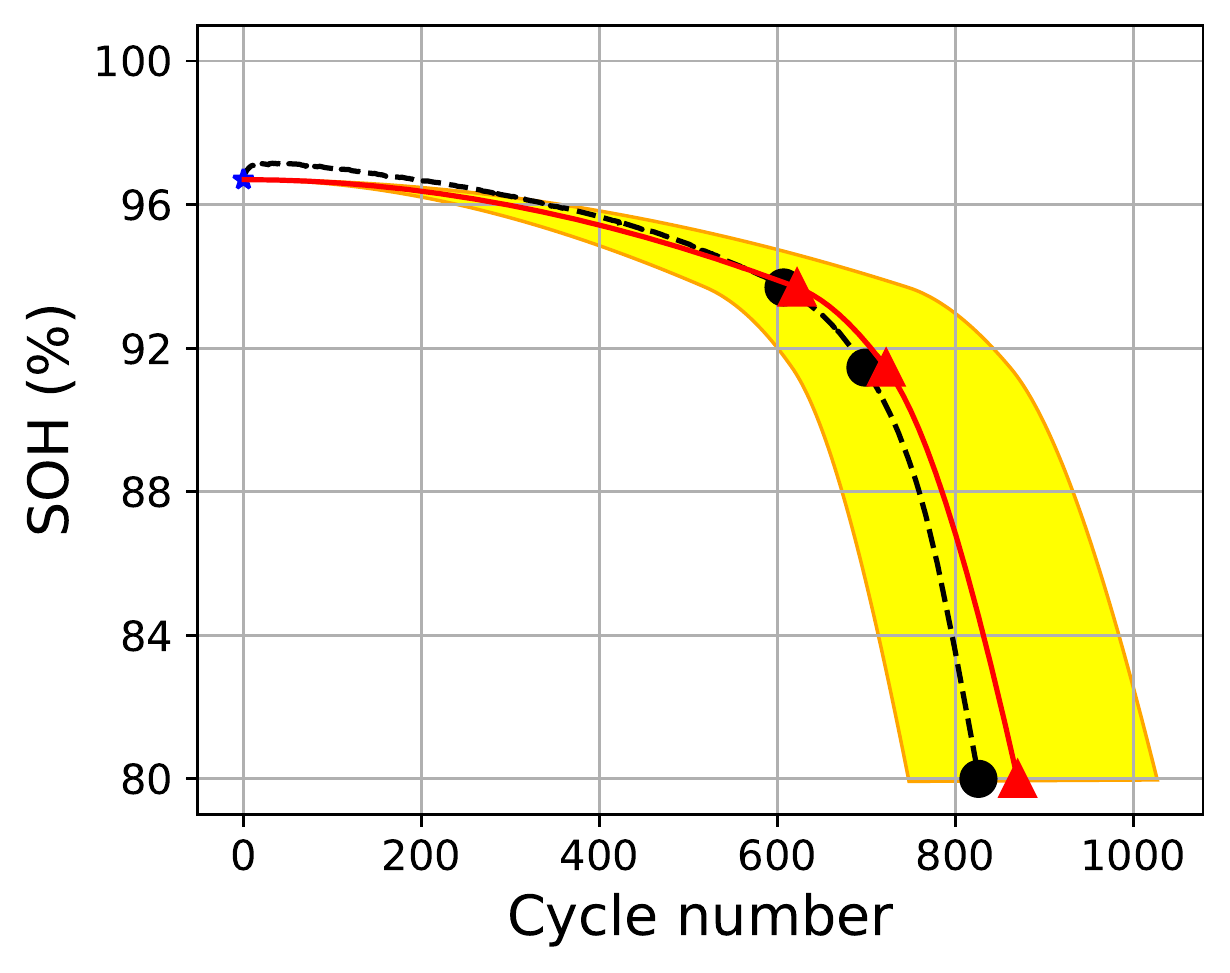}
    }
    \subfigure[]{
    \centering
    \includegraphics[width=0.31\textwidth]{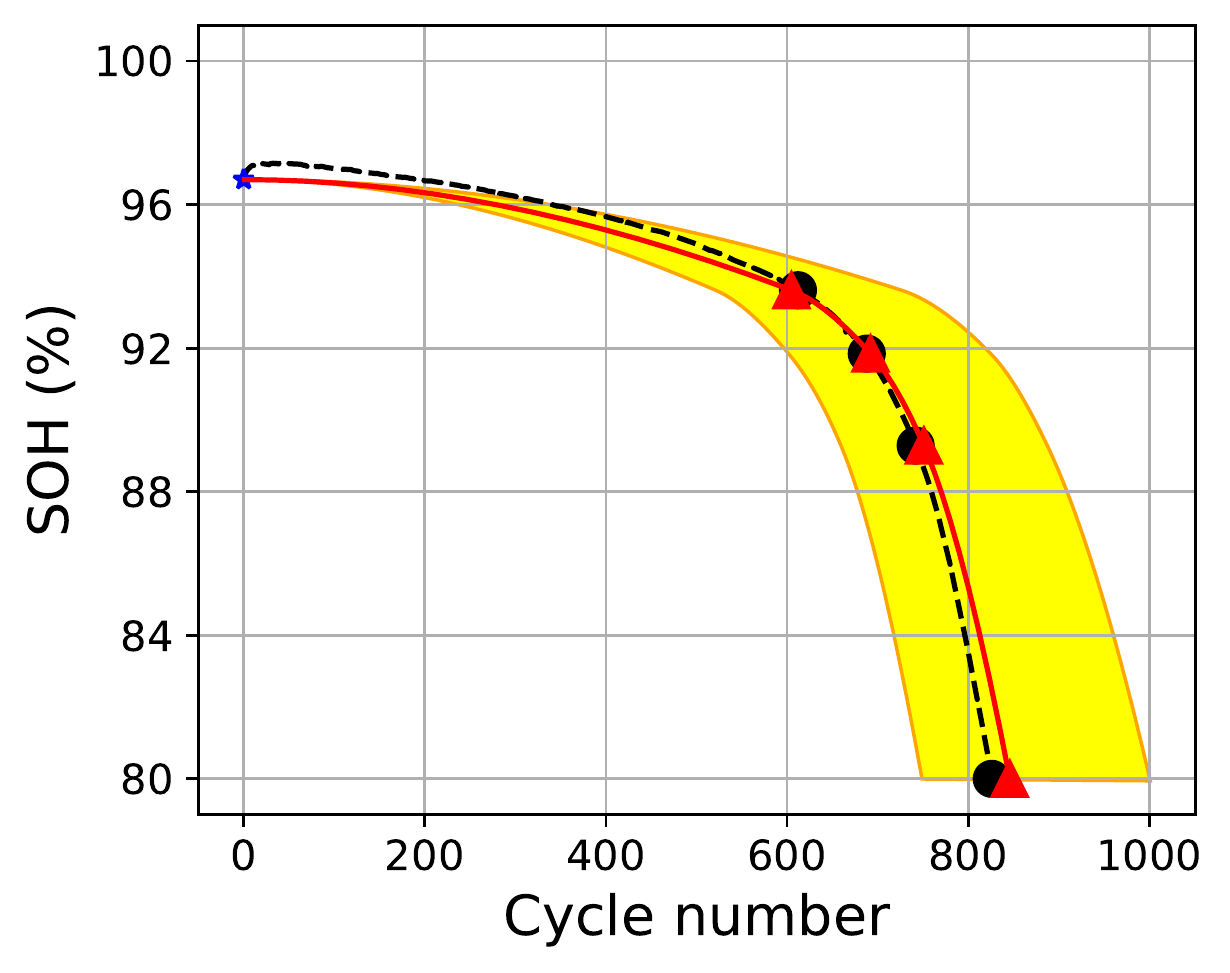}
    }
    \caption{Examples of (a--c) uniform and (d--f) optimized knot-based capacity trajectory prediction with 95\% confidence interval for $K=2,3,4$.}
    \label{fig:Oneshot_prediction}
\end{figure}

Fig.~\ref{fig:Oneshot_prediction} illustrates examples of uniform and optimized knot-based capacity trajectory prediction with a 95\% CI for a cell with $K=2,3,$ and $4$.
\red{
Note that as described in Section~\ref{sec2.2:modelstructure}, the 95\% CIs are obtained using the MC dropout~\cite{gal2015bayesian}.}
The MAEs (MAPEs) in the capacity trajectory on the uniform knots of the cell for $K=2,3,$ and $4$ are 0.0090 Ah (0.97\%), 0.0047 Ah (0.51\%), and 0.0066 Ah (0.72\%), respectively, indicating that two uniform knots are inadequate.
In contrast, the MAEs (MAPEs) in the capacity trajectory on the optimized knots of the cell for $K=2,3,$ and $4$ are 0.0056 Ah (0.62\%), 0.0057 Ah (0.64\%), and 0.0042 Ah (0.46\%), respectively, which shows no significant difference in the number of knots.
The estimated 95\% CI of EOL was investigated according to the number of knots.
The mean CI lengths of all test cells were 80, 88, and 88 cycles for $K=2,3,$ and $4$, respectively.
It is assumed that the CI length increases with the $K$ because the model's prediction error may accumulate, and this also results in an increase in the MAE in knots.
Consequently, the proposed model can directly estimate the overall capacity degradation trajectory by predicting only the cycle numbers of at least two knots based on early single-cycle charge and discharge data.
Thus, it was possible to predict even with a single cycle because all cell data were fully charged and discharged under the controlled temperature and charge-discharge protocols.
In addition, this study used one cycle data containing both charging and discharging, and thus, the features such as Coulomb efficiency (CE) can be implicitly considered to increase the model performance.
%
\red{
Although the proposed method targets the diagnosis of batteries using an early cycle, in order to show the scalability of the proposed method, further experiments are conducted wherein the trained CNN is applied to used batteries (e.g., initial, 100th, 300th, and 500th cycles) using uniform knots, which is described in \ref{appendix:used}.
The results indicate that while the accuracy of the proposed method decreases as the input battery ages, it is possible to predict the capacity degradation pattern even for battery inputs that are not in the initial cycle.}

\subsection{Analysis of input cycle numbers}

To analyze the impact of input cycle numbers on the capacity trajectory prediction, additional experiments were conducted according to total cycle numbers of the input cycle.
Furthermore, a perturbation approach was implemented on the input data in order to evaluate the robustness of the models for different input cycle numbers.
Perturbations were applied by adding Gaussian noise $\Delta X$, with zero mean and variance $\sigma^2$, to the measured input data $X$, in accordance with the suggestions reported in \cite{agarwal1995application}.
The perturbed input data $X'$ can be presented as follows:
\begin{align}
    X'=X+\Delta X,~\text{where}~\Delta X\sim N\left(0, \sigma^2Var(X)\right),
\end{align}
where $Var(X)$ denotes variance of $X$.
The capacity trajectory was predicted using the proposed model according to the perturbed input data, and the MAE deviations between the predicted trajectory with and without input noise were analyzed.
It is worth noting that the voltage, current, and time were selected as input data, and the data were perturbed with $\sigma=0.001, 0.003$, and $0.01$.

\begin{figure}[t]
    \centering
    \includegraphics[width=.75\columnwidth]{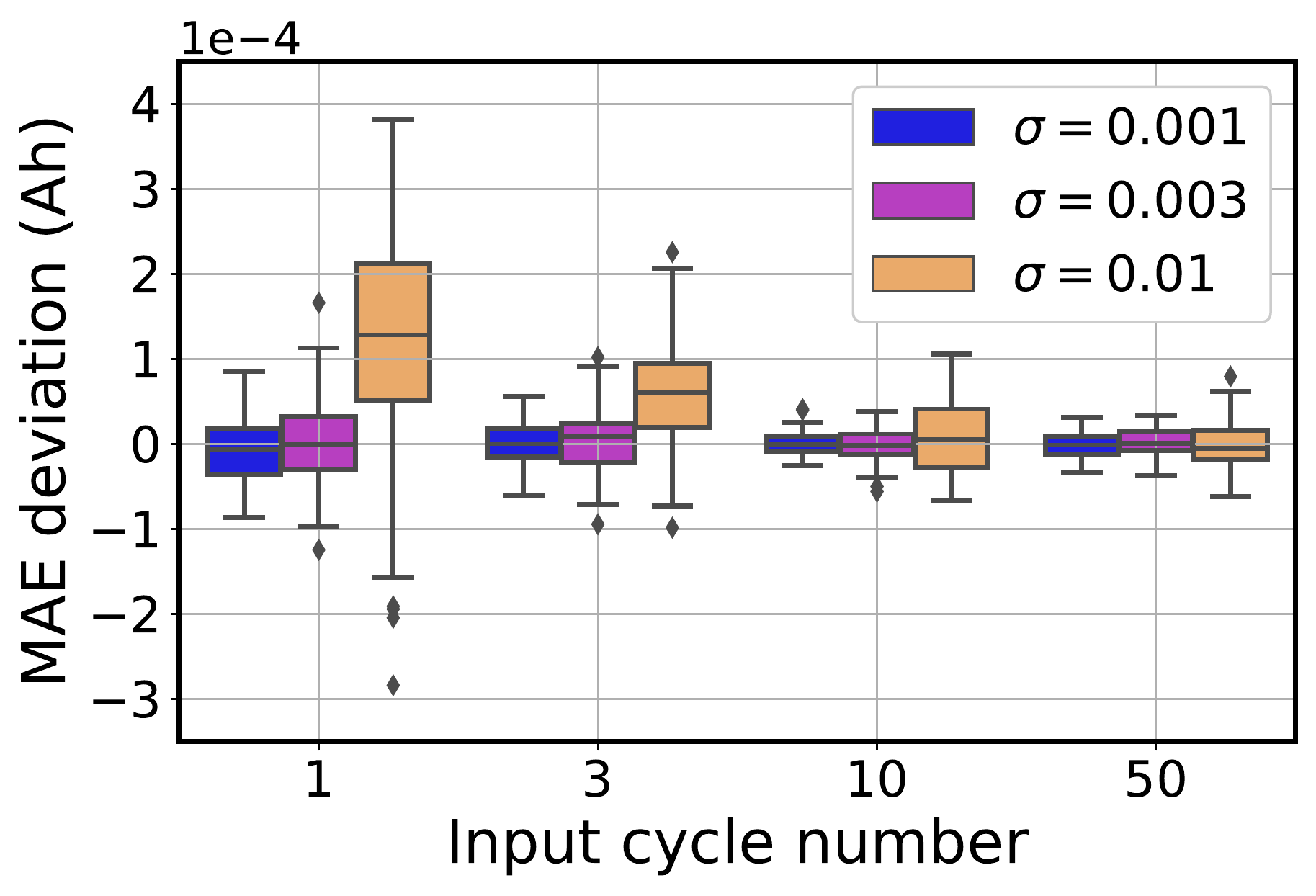}
    \caption{Deviations of MAE in capacity trajectory predictions for different input cycle numbers and $\sigma$s.}
    \label{fig:robustness}
\end{figure}

Table~\ref{tab:robustness} lists the capacity trajectory prediction results of the models for input cycle numbers 1, 3, 10, and 50.
Compared to the case when only one cycle is used, the knot and trajectory predictions do not show significant error reductions according to the number of cycles.
The voltage and current do not vary during the first 50 cycles, and thus, increasing the input cycle numbers may not help improve model accuracy because similar data are additionally received.
\red{
Fig.~\ref{fig:robustness} summarizes the MAE deviations in the trajectory predictions for the 100 perturbed input data for different $\sigma$s and input cycle numbers in the forms of box plots from the evaluation of the robustness of the models under input noise.
The MAE deviation represents a difference of MAEs in the capacity trajectory between using the perturbed and original inputs.}
In the box plot, the central line indicates the median, the edges of the box show the 25th and 75th percentile values, the whiskers extend to the most extreme data points that are not considered outliers, and the outliers are plotted individually and are denoted with `$\blacklozenge$'.
In terms of input cycles, as the noise level increases, the range of the deviations tends to increase.
If a small cycle is used, the median of the device according to the noise level is also increased.
However, the MAE deviations according to the noise level of input data show significant changes for different input cycle numbers.
As the number of input cycles increases, the increase in both median and range reduces.
These results indicate that the input cycle number does not have a significant impact on the predictive accuracy of the trajectory, while the robustness under the noise of the input data increases as the input cycle numbers increase.
In general, the observation error of $\sigma=0.01$ does not occur frequently, so using a three-cycle input is recommended to obtain both robustness and efficiency.

\begin{table}[b]
\centering
\caption{MAE and MAPE in optimized knot prediction and capacity trajectory prediction using the proposed model for input cycle numbers 1, 3, 10, and 50.}
\begin{tabular}{cccccc}
\toprule
\multirow{2}{*}{} & \multirow{2}{*}{Input cycle numbers} & \multicolumn{2}{c}{Knot prediction} & \multicolumn{2}{c}{Trajectory prediction}\\
\cmidrule{3-6}
&& MAE (Cycle) & MAPE (\%) & MAE (Ah) & MAPE (\%)\\
\midrule
& 1   & 75 & 10.97 & 0.0133 & 1.33 \\
& 3   & 69 & 10.44 & 0.0122 & 1.22 \\
& 10   & 78 & 11.79 & 0.0122 & 1.23 \\
& 50   & 77 & 10.79 & 0.0121 & 1.22 \\
\bottomrule
\end{tabular}
\label{tab:robustness}
\end{table}

\subsection{Capacity trajectory prediction for various patterns}

\begin{figure}[t]
    \centering
    \subfigure[]{
    \centering
    \includegraphics[width=0.31\textwidth]{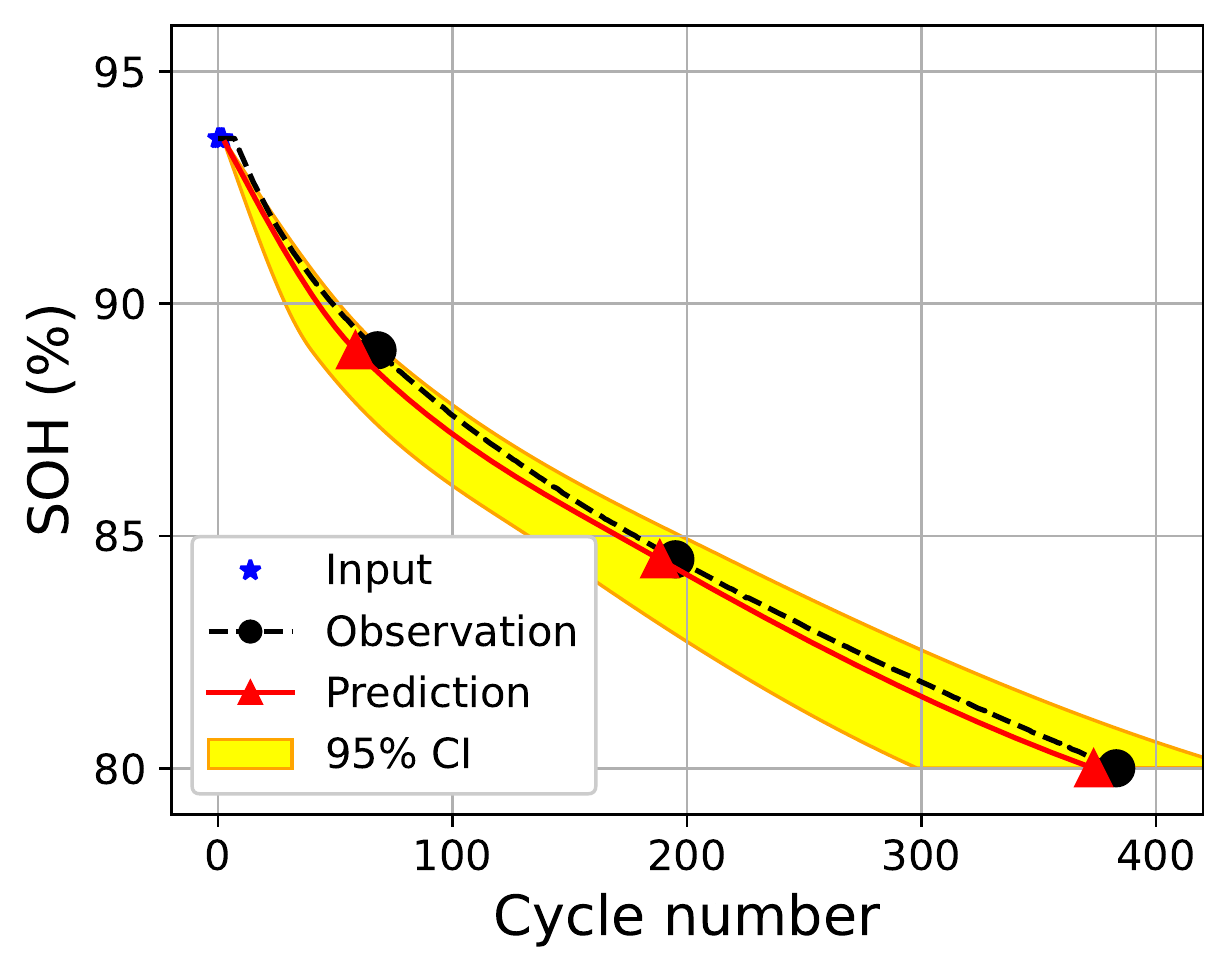}
    }
    \subfigure[]{
    \centering
    \includegraphics[width=0.31\textwidth]{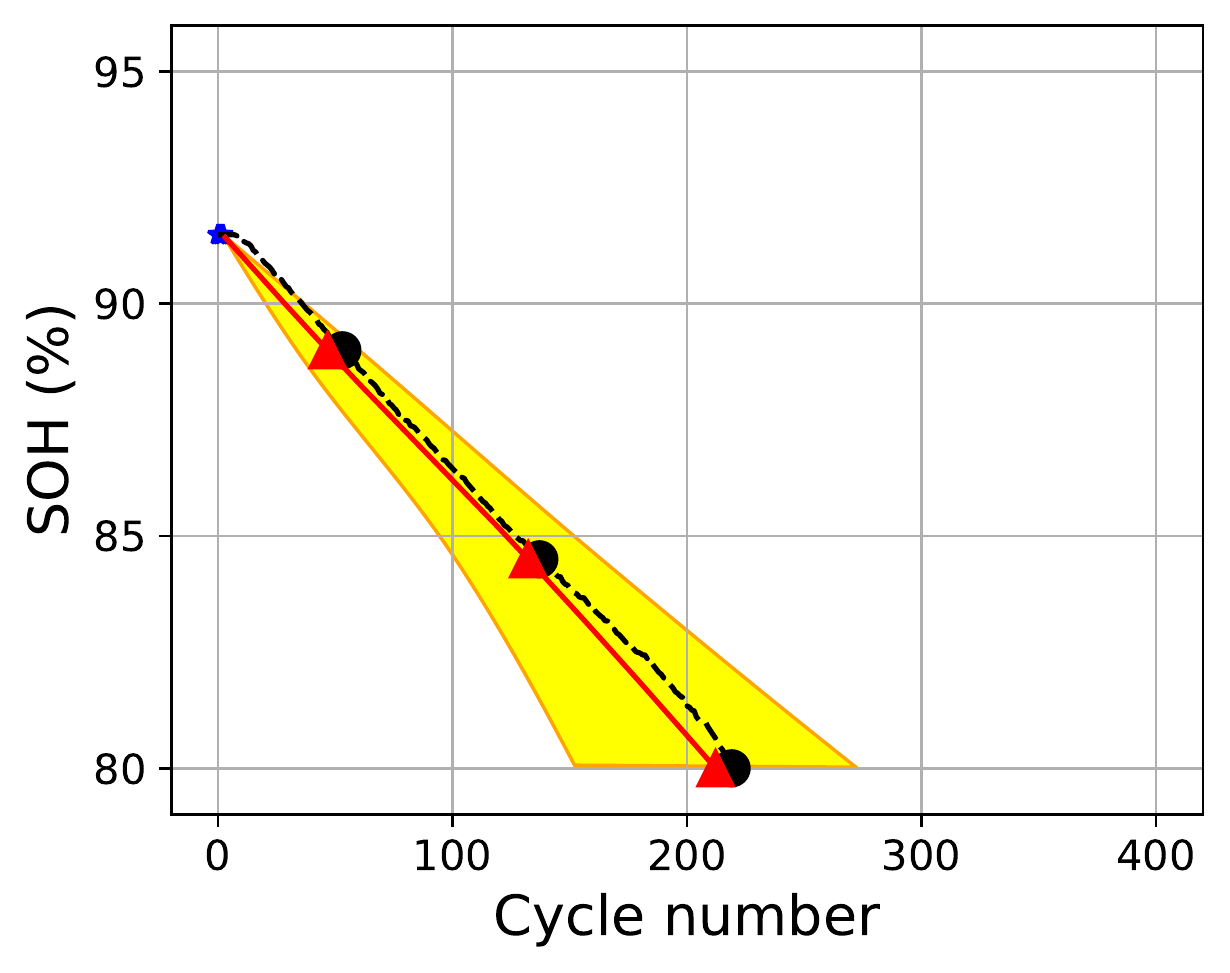}
    }
    \subfigure[]{
    \centering
    \includegraphics[width=0.31\textwidth]{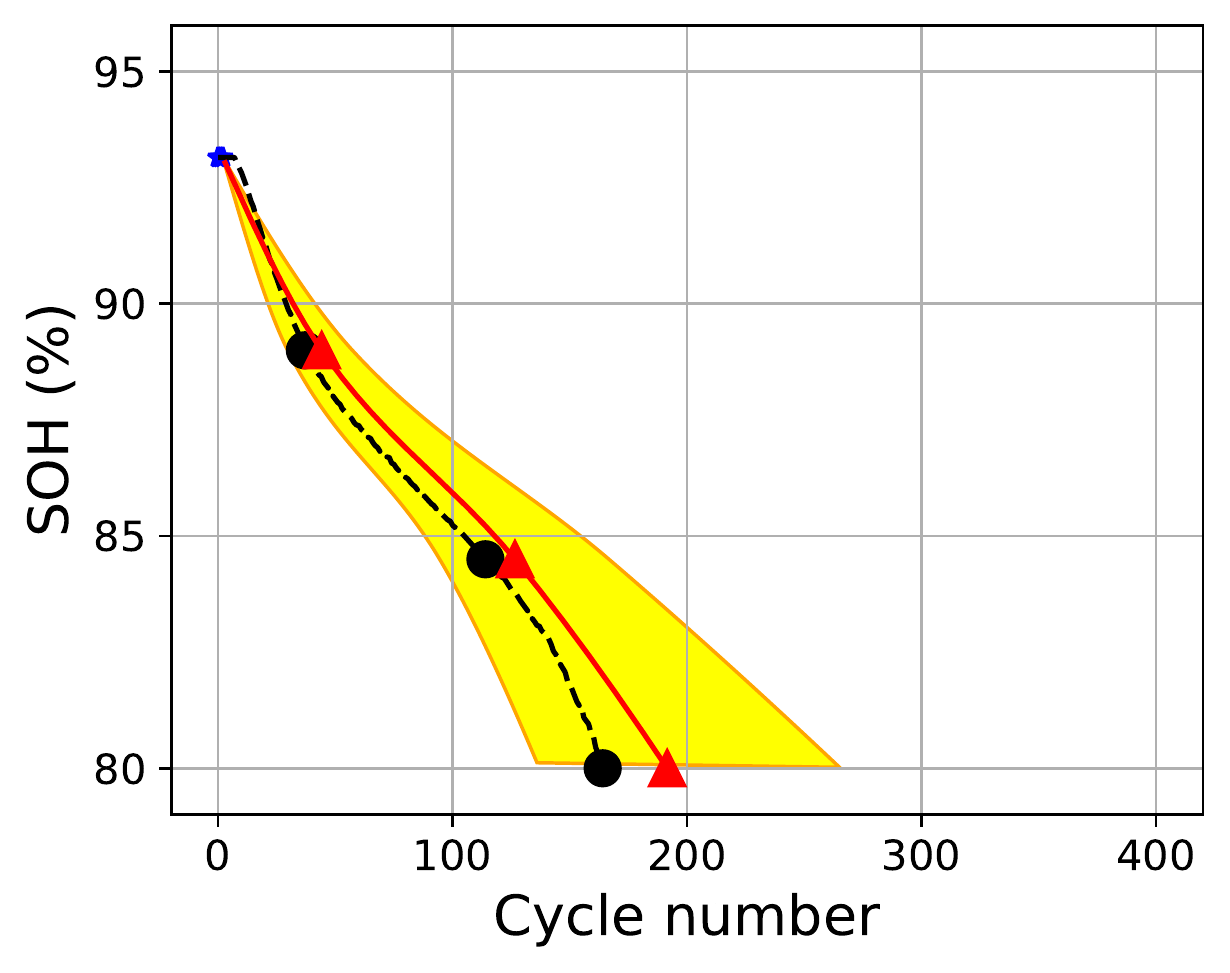}
    }
    \caption{Examples of capacity trajectory prediction with $\pm1\sigma$ confidence interval for three different aging patterns~\cite{zhu2022data}.}
    \label{fig:Oneshot_prediction_tongji}
\end{figure}

We demonstrate the proposed capacity trajectory prediction method using Dataset 2, NCA and NCM cell data with various degradation patterns~\cite{zhu2022data}.
Considering the results in Tables~\ref{tab:knotprediction_results} and \ref{tab:robustness}, three uniform knots and three-cycle input were used in the proposed model.
Because Dataset 2 contains two different chemistry types in 82 cells, stratified five-fold cross-validation was applied, which preserves the percentage of samples for each chemistry.
Fig.~\ref{fig:Oneshot_prediction_tongji} illustrates examples of prediction results with CI for three degradation patterns: (a) sublinear, (b) linear, and (c) knee patterns.
Note that the sublinear degradation pattern is attributed to solid-electrolyte interphase (SEI) growth, and the linear degradation pattern is attributed to active material loss~\cite{attia2022knees}.
The MAEs (MAPEs) of the predicted knots and capacity trajectory are 22 cycles (17.73\%) and 0.0106 Ah (1.24\%), respectively, which are similar to those in Dataset 1.
In addition, CIs of three examples encompass the observed capacity trajectory.
The CI length increases with cycle numbers due to the accumulation of model prediction errors, as shown in Fig.~\ref{fig:Oneshot_prediction_tongji}.
This result indicates that for general applicability, the proposed model-free reconstruction approach can be applied to represent the capacity degradation regardless of the shapes of the trajectories.

\section{Conclusions}\label{sec4:conclusion}

This study proposes a knot-based capacity trajectory prediction for LIB cells using a CNN model.
To predict the full-capacity trajectory, 2--4 knots were identified according to the uniform division or Bayesian optimization, and the PCHIP interpolation was applied to construct the trajectory on the knots.
The proposed method was validated through a dataset consisting of 169 LFP cells using five-fold cross-validation, and it was observed that almost all the MAEs and MAPEs in trajectory prediction were less than 0.0150 Ah and 1.60\%, respectively.
The results indicate that the proposed model can directly estimate the overall capacity degradation trajectory by predicting only the cycle numbers of at least two knots using early single-cycle charge and discharge data.
\red{
Furthermore, the comparison between the uniform and optimized knots shows that uniform knots sufficiently perform better compared to optimized knots in capacity degradation trajectory prediction, which can be an important finding of interest to future researchers in the field.}
In the robustness evaluation of the proposed model, it is recommended to use a three-cycle input to obtain both robustness and efficiency, considering the observational noise of the input data.
To demonstrate the applicability on various capacity degradation, the proposed method was applied to 82 NCA and NCM cells containing sublinear, linear, and knee patterns, where the MAE and MAPE in trajectory prediction indicate similar or better performances of the method to those in LFP cells.
\red{
These results indicate that the proposed model-free reconstruction method has the flexibility to predict various degradation patterns.
Although the assumption of the monotonic decrease in the battery capacity does not help capture characteristics such as temporary capacity regeneration, it provides inductive bias for better generalization of the prediction model in predicting the overall capacity trajectory.}
In the battery manufacturing and diagnosis processes, quantitative predictions, including confidence intervals, can help establish appropriate warranties or replacement cycles.


\section*{Acknowledgment}

%
%
This work was supported by the Korea Institute of Energy Technology Evaluation and Planning (KETEP) grants funded by the Ministry of Trade, Industry \& Energy, Republic of Korea (No.20214910100070).
Computing resources were supported by the National Supercomputing Center (KSC-2022-CRE-0131) and the National IT Industry Promotion Agency (NIPA).
We also would like to thank anonymous reviewers for constructive comments and careful reading of the manuscript.

\bibliographystyle{elsarticle-num}
\bibliography{bibfile}

\appendix

\section{\red{Bayesian optimization for knot optimization}}\label{appendix:bo}

\red{
Bayesian optimization is an adaptive search algorithm that builds an agent model according to the Gaussian process and can provide the most appropriate SOH levels for knots.
Given the number of knots $K$, the objective of Bayesian optimization is to determine the SOH levels of knots $\mathbf{p} = [p_1, p_2, \ldots, p_K]^T$ to minimize a function $d(\mathbf{p})$, which represents an interpolation error between the observed and reconstructed capacity trajectory as follows:
\begin{align}
    d(\mathbf{p}) = \frac{1}{n}\sum_{i=1}^n \left|y_i - \tilde{y}_{i}(\mathbf{p})\right|,\label{eq:BO_loss}
\end{align}
where $i$ refers to the cycle number, $y$ is the observed capacity trajectory, and $\tilde{y}$ is the constructed capacity trajectory using PCHIP interpolation.
To achieve the global optimum of $\mathbf{p}$ in Bayesian optimization, a Gaussian process (GP) is widely used as a probabilistic surrogate model of the function $d(\mathbf{p})$, and acquisition functions are constructed based on the surrogate model to determine the next optimal experimental points by balancing the trade-off between exploitation and exploration~\cite{lookman2019active, gramacy2020surrogates}.
Given $M$ sampled points $\mathcal{P}=\left\{\mathbf{p}_m\right\}_{m=1}^M$, the expected improvement (EI) is a widely used acquisition function, defined as follows.
\begin{align}
    EI_M(\mathbf{p}) = \begin{cases}
        \left(d(\mathbf{p}^+) - \mu(\mathbf{p}) - \zeta\right) \Phi(Z) + \sigma(\mathbf{p}) \phi (Z),~&\text{if}~\sigma(\mathbf{p}) > 0\\
        0,~&\text{if}~\sigma(\mathbf{p}) = 0
    \end{cases},
\end{align}
where $\mathbf{p}^+=\underset{\mathbf{p}_m\in \mathcal P}{\arg\!\min}~d(\mathbf{p}_m)$; $\mu(\mathbf{p})$ and $\sigma(\mathbf{p})$ are the posterior predictive marginal GP mean and standard deviation, respectively; $\zeta \ge 0$ is an adjustable margin parameter;  $Z=\left(d(\mathbf{p}^+) - u(\mathbf{p}) - \zeta)\right)/\sigma(\mathbf{p})$ if $\sigma(\mathbf{p}) > 0$ and $Z=0$ otherwise; and $\phi(\cdot)$ and $\Phi(\cdot)$ are the Gaussian probability density function and cumulative distribution function, respectively.
The Bayesian optimization is then iteratively evaluated at the point with the largest EI as follows:
\begin{align}
    \mathbf{p}_{M+1}=\underset{\mathbf{p}}{\arg\!\max}~{EI_M(\mathbf{p})}.\label{eq:BO_final}
\end{align}
The selected knots $\mathbf{p}_{M+1}$ are then added to $\mathcal{P}$.
After sufficient iterations of Bayesian optimization, the optimal knot $\mathbf{p}$ is obtained.
Subsequently, the capacity trajectory can be obtained by precisely predicting the cycle numbers of the optimized knots.
}

\section{Case study for selecting input variables}\label{appendix:tuning}

The predictive performance varied depending on the input variables when implemented using experimental datasets and algorithms.
Thus, a case study was conducted to determine the optimal case through experiments on various  combinations of input variables, including voltage, current, time, capacity, and temperature.
Voltage was added essentially, and current and time were selected together, or only capacity was selected because capacity is a variable wherein current and time are integrated.
Temperature was included with or without other variables.
Thus, four cases were compared, as shown in the Table~\ref{tab:Input_case}.

\begin{table}[t]
    \renewcommand{\thetable}{B.1}
    \centering
    \caption{Cases of Input variables}
    \scalebox{1}{
    \begin{tabular}{ccccccc}
         \toprule
         Case & Voltage & Current & Time & Capacity & Temperature \\
         \midrule
         1 & O & O & O & X & X \\
         2 & O & X & X & O & X \\
         3 & O & O & O & X & O \\
         4 & O & X & X & O & O \\
         \bottomrule
     \end{tabular}
    }
    \label{tab:Input_case}
\end{table}

\begin{figure}[t]
    \renewcommand{\thefigure}{B.1}
    \centering
    \subfigure[]{
    \includegraphics[width=.45\columnwidth]{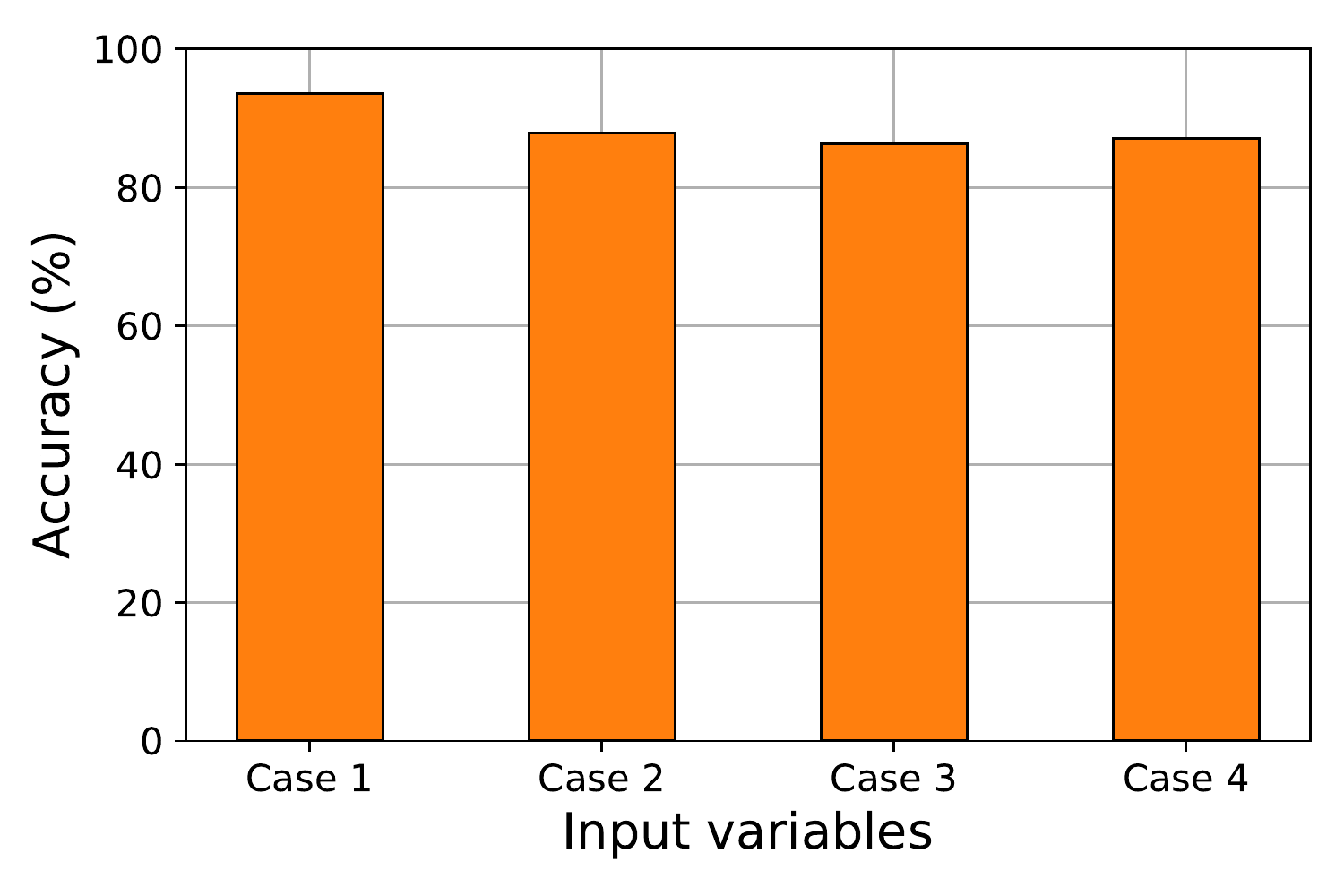}
    }
    \subfigure[]{
    \includegraphics[width=.45\columnwidth]{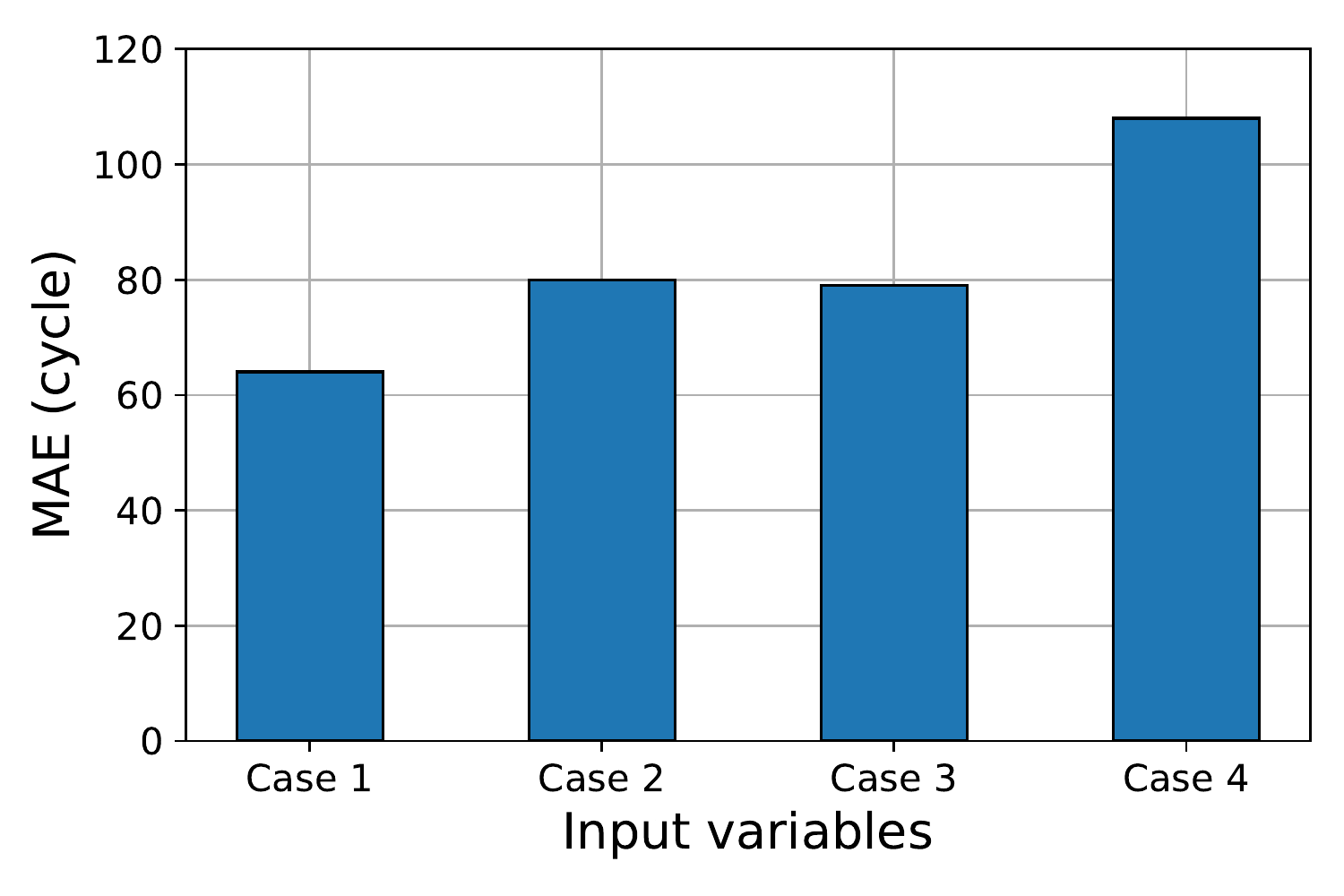}
    }
    \caption{
    Knee point classification and prediction results with different input variables
    }
    \label{fig:case_prediction}
\end{figure}

A case study of the input parameters for the knee point classification was first conducted, and the result is shown in Fig.~\ref{fig:case_prediction}~(a).
Case 1, a set of voltage, current, and time, was selected as the best case because it exhibited the best accuracy of 93.6\%.
For a clearer understanding, when temperature was not included in the input (cases 1 and 2), accuracy was greater than that of inclusion (cases 3 and 4), while the accuracy decreased when capacity was used (cases 2 and 4).
Thus, it is assumed that temperature hinders the prediction of the life of the battery, and it is better to use current and time rather than capacity alone as input.
The case study for the knee point prediction is implemented in the same manner as the classification problem, and the result is shown in Fig.~\ref{fig:case_prediction}~(b).
Similar to the classification problem, case 1 was selected as the best case with the lowest MAE.
The MAE of the knee point prediction in the first case achieved 64 cycles, which was the best result among the four cases.
Consequently, voltage, current, and time were used as inputs in this study.

\section{\red{Knee point classification and prediction}}\label{sec:kneepoint_prediction}

\red{
To evaluate the feature extraction performance of the CNN model, knee point classification and prediction were validated.
It is worth noting that for a fair comparison, only the experimental data from Severson \textit{et al}.~\cite{severson2019data} were used, as by Ferm{\'\i}n-Cueto \textit{et al}.~\cite{fermin2020identification}.
Following Ferm{\'\i}n-Cueto \textit{et al}.~\cite{fermin2020identification}, the Bacon and Watts (BW) model~\cite{bacon1971estimating} was used to identify the knee points, which is formulated as follows.
\begin{align}
    Y=\alpha_0 + \alpha_1(x- x_1) + \alpha_2(x-x_1)\tanh\{(x-x_1)/\gamma\} + Z,
\end{align}
where $Z$ is a normally distributed and zero-mean residuals, $\alpha_1$ and $\alpha_2$ are the slopes of the intersecting lines, $\alpha_0$ is an intercept of the leftmost segment, and $\gamma$ is the abruptness of the transition.
The parameters of the BW model were identified using Levenberg–Marquardt nonlinear least squares algorithm, implemented in \texttt{curve\_fit} function of \texttt{SciPy}~\cite{2020SciPy-NMeth}.
}

\red{
The shapes of the FC layers vary depending on the model's goals (e.g., knee point classification, knee point prediction, and capacity degradation trajectory prediction).
The primary aim of knee point classification is to classify the knee points of the capacity degradation curves of battery cells, where the knee points are divided into three classes according to the criteria used by Ferm{\'\i}n-Cueto \textit{et al}.~\cite{fermin2020identification} as follows:
\begin{align}
    0 \le C_{1} < 500 \le C_{2} < 1100 \le C_{3}, \label{class_cri}
\end{align}
where $C_{1}$, $C_{2}$, and $C_{3}$ are short, medium, and long classes, respectively.
The classes were further divided according to this criterion to compare the performance of the proposed model with the support vector machine (SVM) used in the previous study~\cite{fermin2020identification}.
Denoting the CNN model for knee point classification as $f_{cl}$, the convolution layer receives data $\mathbf{X}$ as input, the FC layer outputs three-dimensional real values $\mathbf{z}=[z_{1}, z_{2}, z_{3}]^T$, and, finally, the softmax function is applied to obtain the probability of belonging to the corresponding class.
\begin{align}
    f_{cl}(\mathbf{X}) = \begin{bmatrix}
        p(C_1 | \mathbf{X}) \\ p(C_2 | \mathbf{X}) \\ p(C_3 | \mathbf{X})
    \end{bmatrix} = \frac{1}{\sum_{k=1}^3\exp{(z_{k})}}\begin{bmatrix}
        \exp (z_{1}) \\ \exp (z_{2}) \\ \exp (z_{3})
    \end{bmatrix},
\end{align}
where $p(C_k | \mathbf{X})$ represents the probability of the knee point belonging to class $C_k$ given input $\mathbf{X}$.
The knee points can be identified using gradient-based methods with the BW model~\cite{fermin2020identification}.
The adim of knee point prediction is to predict the cycle numbers of the knee points of the capacity degradation curves in the battery cells.
Denoting the CNN model for the knee point prediction as $f_{pr}$, the FC layer outputs a one-dimensional real value $z$ because the cycle number values are directly predicted; that is, $f_{pr}(\mathbf{X})=z$.
}

\begin{figure}[t]
    \renewcommand{\thefigure}{C.1}
    \centering
    \subfigure[]{
    \centering
    \includegraphics[width=0.28\textwidth]{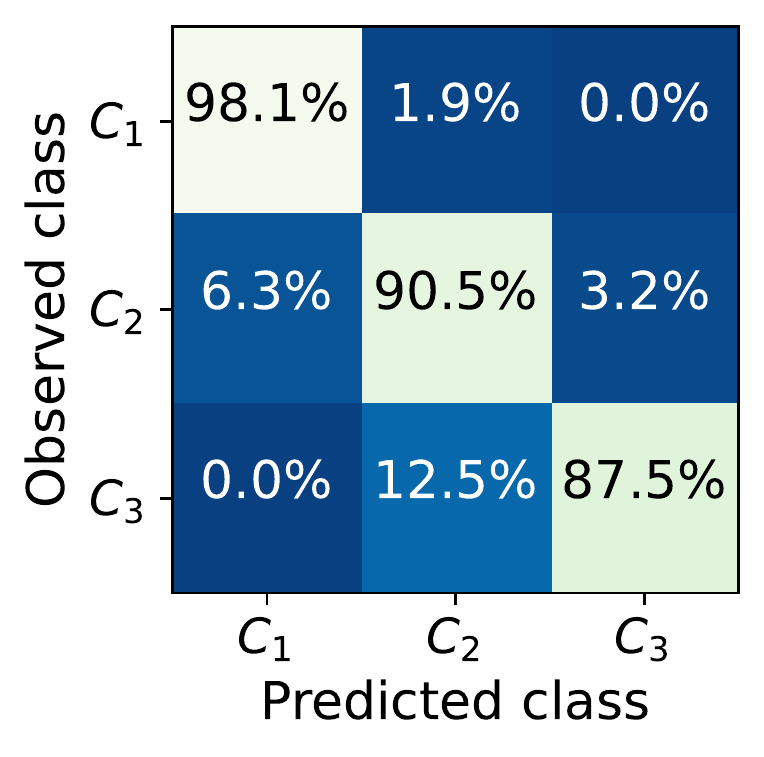}
    }
    \subfigure[]{
    \centering
    \includegraphics[width=0.65\textwidth]{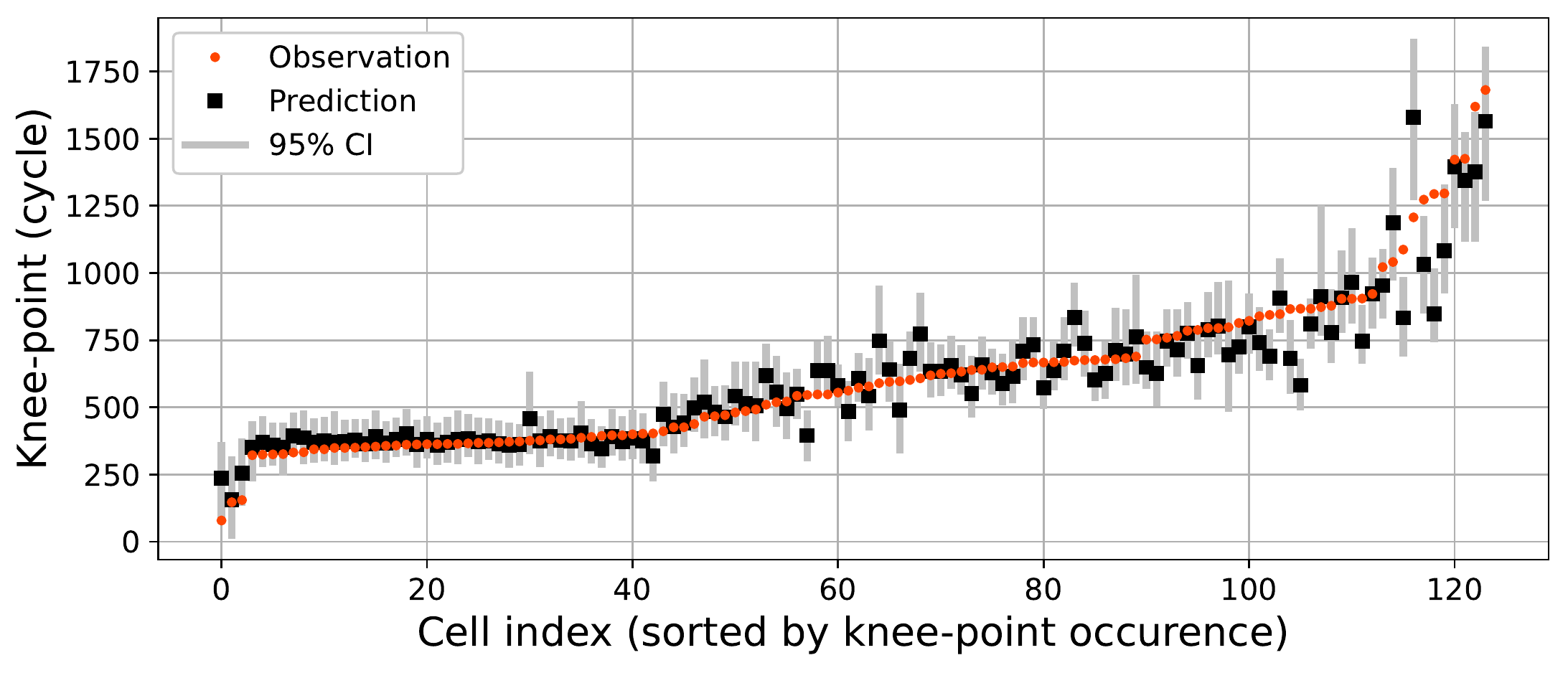}
    }
    \caption{(a) Confusion matrices of knee point classification and (b) results of knee point prediction with 95\% confidence intervals for 124 cells using the first single cycles.}
    \label{fig:knee_pred_result}
\end{figure}

\red{
Fig.~\ref{fig:knee_pred_result} (a) shows confusion matrices showing the result of classifying knee points for the proposed method, where the overall accuracy was 93.6\%.
For the short and medium classes, the results were 98.1\% and 90.5\%, respectively.
The accuracy was 87.5\% when the observed values of knee points were long because the data distribution was primarily concentrated in the medium class and rarely concentrated in the long class of knee points.
As stated in Ferm{\'\i}n-Cueto \textit{et al}.~\cite{fermin2020identification}, the most common errors are those that underestimate the lifespan of the cell, which is preferable because the model allows conservative evaluation.
Most importantly, the accuracy for the short-observed class was remarkably high, implying that the model can be prepared for cases with fast capacity degradation.
For a detailed comparison, the method in Ferm{\'\i}n-Cueto \textit{et al}.~\cite{fermin2020identification} showed classification results of 92.5\%, 84.1\%, and 87.5\% for three classes, where the overall accuracy was 88.0\% using three cycles as the input.
It was confirmed that our proposed method shows greater accuracy for short and medium observed classes, while showing identical results in the long class.
The method in Strange and Dos Reis~\cite{strange2021prediction} yielded  classification results of 92.8\%, 100.0\%, and 100.0\% for three classes using a single cycle as the input.
However, a direct performance comparison is difficult because the simulation was performed based on a single train-test split instead of cross-validation or leave-one-out validation.
Consequently, we have shown that the classification of knee points can be predicted with high accuracy using data obtained through the initial first-cycle charge and discharge of batteries using a CNN.
}

\red{
%
Fig.~\ref{fig:knee_pred_result}~(b) illustrates the results of predicting the knee point cycle numbers with 95\% confidence intervals for 124 cells using the initial single-cycle data.
The confidence intervals were obtained using MC dropout, as stated in Section~\ref{sec2.2:modelstructure}.
The proposed method predicts the knee point cycle numbers with an MAE and MAPE of 64 cycles and 11.1\%, respectively.
For regions $C_1$, $C_2$, and $C_3$, wherein the observed cycle numbers of the knee points are divided into three regions, the MAEs (MAPEs) are 29 (11.50\%), 73 (10.17\%), and 217 (16.26\%) cycles, respectively.
Similar to the knee point classification, the MAE is relatively large in region $C_3$ due to the data imbalance problem, where most batteries have knee points in $C_1$ and $C_2$.
However, because the observed cycle numbers were smaller in $C_1$ than in $C_2$, the MAPE in $C_1$ was larger than in $C_2$, in contrast to the MAE.
In addition, for a comparison of the results of knee point predictions, the MAE and MAPE in knee point predictions and corresponding results reported in previous studies are given in Table~\ref{tab:kneepoint_comparison}.
In the case of Strange and Dos Reis~\cite{strange2021prediction}, wherein only one input cycle used as in the proposed method, the MAE and MAPE are the smallest, and the proposed method shows similar performance.
Notably, a direct performance comparison was difficult because the simulation was performed based on a single train-test split.
Ferm{\'\i}n-Cueto \textit{et al}.~\cite{fermin2020identification} also achieved great performance in knee point predictions using the first 50 cycles.
They extracted several features from cycle data, and Table~\ref{tab:kneepoint_comparison} ensures that the proposed automatic feature selection using the CNN can perform better predictions that the manual feature selection.
Saxena \textit{et al}.~\cite{saxena2022convolutional} used the first 100 cycles, but did not achieve better performance than the others.
This might be due to a single train-test split, which cannot provide the robust prediction results.
Nevertheless, we confirmed that the CNN model can sufficiently classify and predict knee points from the initial cycle data without manual feature selection.
}

\begin{table}[t]
\renewcommand{\thetable}{C.1}
\centering
\caption{Mean absolute error (MAE) and mean absolute percentage error (MAPE) for knee point predictions using the proposed method. Corresponding results reported in previous studies are provided for comparison.}
\begin{tabular}{lccc}
\toprule
& Number of input cycles & MAE (Cycles) & MAPE (\%)\\
\midrule
Present method & 1 & 64 & 11.1 \\
Strange and Dos Reis~\cite{strange2021prediction} & 1 & 55 & 9.7 \\
Ferm{\'\i}n-Cueto \textit{et al}.~\cite{fermin2020identification} & 50 & 58 & 9.4 \\
Saxena \textit{et al}.~\cite{saxena2022convolutional} & 100 & 97 & 15.0 \\
\bottomrule
\end{tabular}
\label{tab:kneepoint_comparison}
\end{table}

\section{\red{Application of proposed method on used batteries}}\label{appendix:used}

\begin{figure}[t]
    \renewcommand{\thefigure}{D.1}
    \centering
    \subfigure[]{
    \centering
    \includegraphics[width=0.31\textwidth]{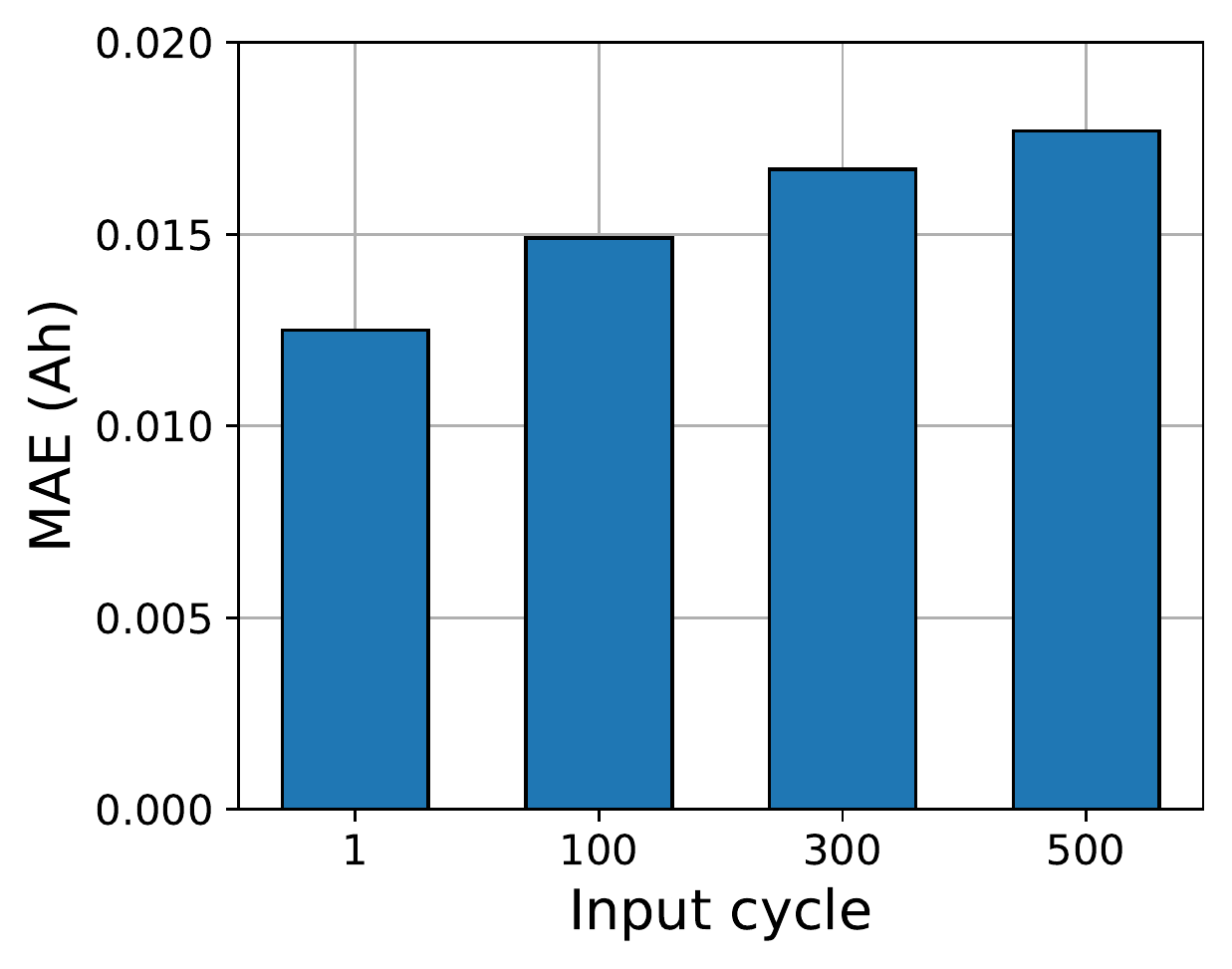}
    }
    \subfigure[]{
    \centering
    \includegraphics[width=0.31\textwidth]{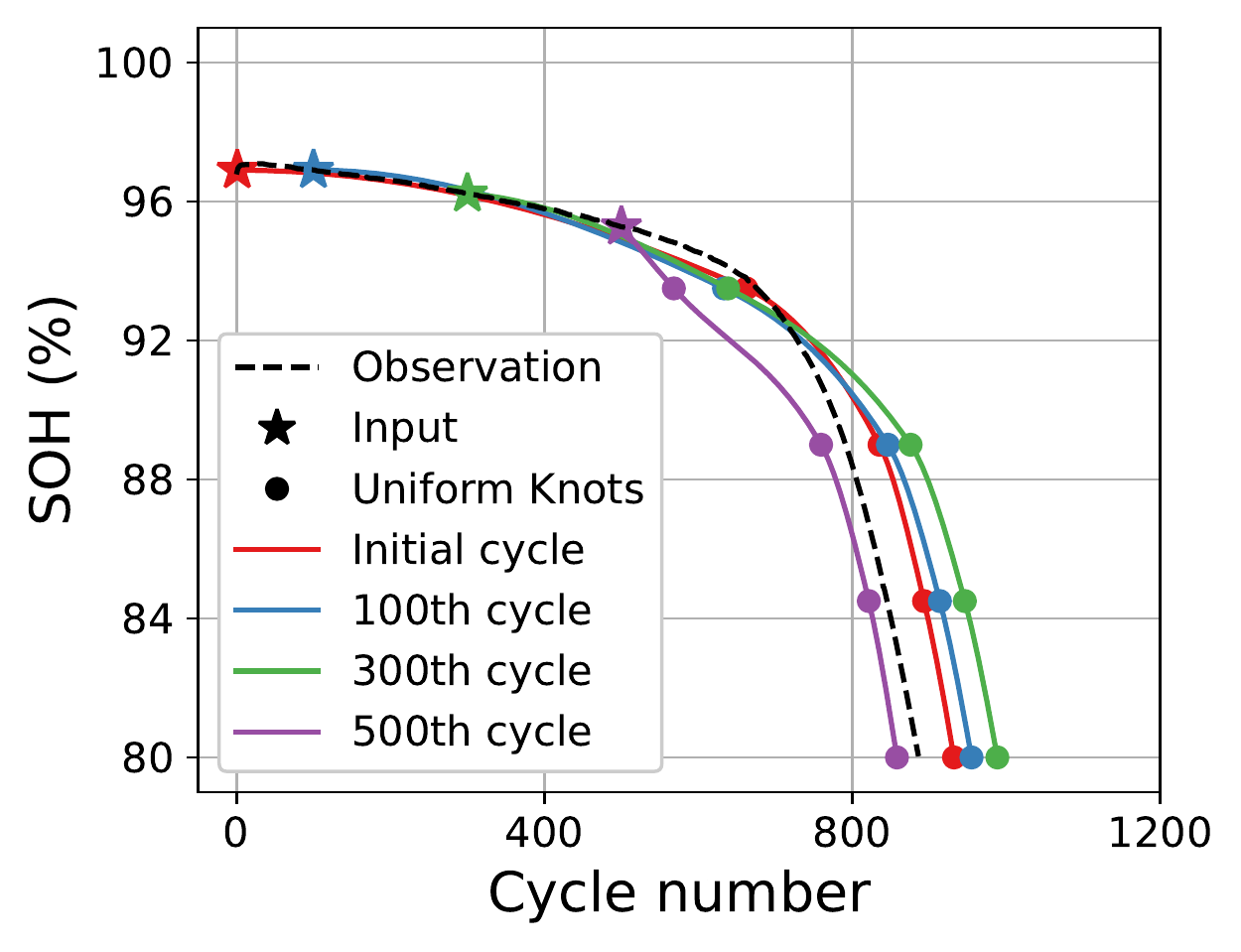}
    }
    \subfigure[]{
    \centering
    \includegraphics[width=0.31\textwidth]{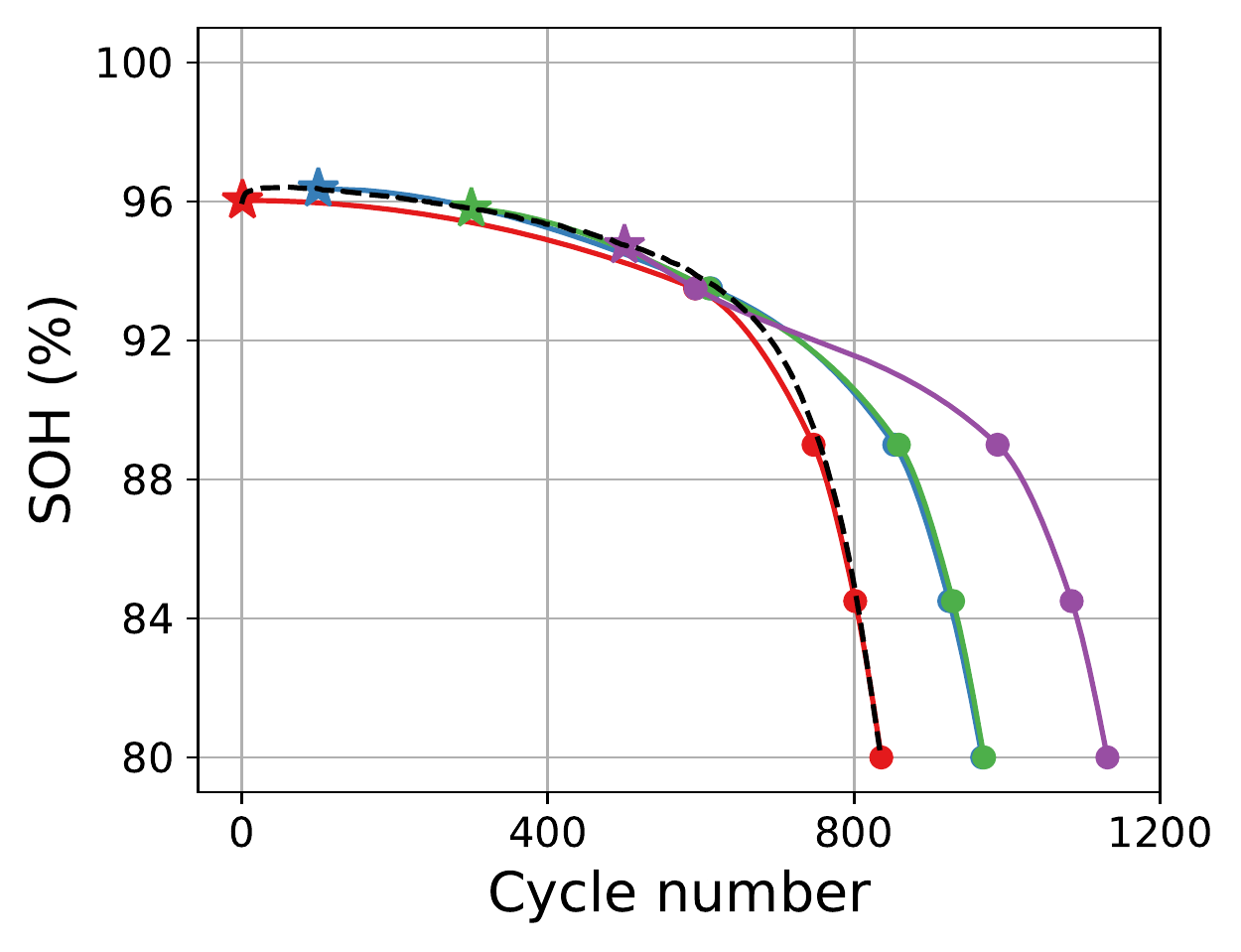}
    }
    \caption{(a) MAEs in capacity trajectory for each input cycle and (b, c) two examples of trajectory prediction.}
    \label{fig:used_battery}
\end{figure}

\red{
Although the proposed method targets the diagnosis of batteries using an early cycle, in order to show the scalability of the proposed method, further experiments are conducted applying the trained CNN on used batteries (e.g., initial, 100th, 300th, and 500th cycles) using uniform knots.
Fig.~\ref{fig:used_battery} illustrates MAEs in capacity trajectory for each input cycle and two examples of trajectory prediction.
As shown in Fig.~\ref{fig:used_battery} (a), the MAEs (MAPEs) of the knots are 0.0125 (1.37\%), 0.0149 (1.60\%), 0.0167 (1.84\%), and 0.0177 Ah (1.95\%) for the initial, 100th, 300th, and 500th cycles, respectively, which indicates the accuracy of the proposed method decreases as the input battery ages.
The best and worst results of trajectory predictions with different input cycles are illustrated in Fig.~\ref{fig:used_battery} (b) and (c), respectively.
In Fig.~\ref{fig:used_battery} (b), the predicted trajectory changes accordingly, even when an aging battery enters.
However, as in Fig.~\ref{fig:used_battery} (c), the predicted trajectory is pushed backward as long as input cycles when an aged battery enters.
The results indicate that while the accuracy of the proposed method decreases for aged batteries, it is possible to predict the capacity degradation pattern even for battery inputs that are not in the initial cycle.}

\end{document}